\documentclass[a4paper,12pt]{article}

\usepackage[utf8]{inputenc}
\usepackage[T1]{fontenc}
\usepackage[swedish,english]{babel}

\usepackage{color}
\usepackage{amssymb}
\usepackage{amsmath}
\usepackage{slashbox}
\usepackage{ae}
\usepackage{slashed}
\usepackage{graphics}
\usepackage{graphicx}
\usepackage{epstopdf}
\usepackage{url}

\newcommand{\rd}{\ensuremath{\mathrm{d}}}
\newcommand{\id}{\ensuremath{\,\rd}}
\newcommand{\g}{\gamma}
\newcommand{\lb}{\bar{\lambda}}
\newcommand{\la}{\lambda}
\newcommand{\La}{\Lambda}
\newcommand{\w}{\wedge}
\newcommand{\e}{\varepsilon}
\newcommand{\lbracket}{[\![}
\newcommand{\rbracket}{]\!]}
\newcommand{\fb}{\bar{f}}
\newcommand{\li}{\hspace{1mm}}

\usepackage{cite}

\usepackage{epsfig}
\usepackage{times}
\usepackage[small]{caption}
\DeclareMathOperator{\tr}{tr} 
\usepackage{ifthen}
\usepackage{longtable}
\usepackage{multirow}
\numberwithin{equation}{section} 
\usepackage{fancyhdr}
\begin{document}
\selectlanguage{english}
\setcounter{secnumdepth}{3}
\frenchspacing               
\pagenumbering{arabic}
\null{\vspace{\stretch{1}}}
\includegraphics*[width=1.8cm]{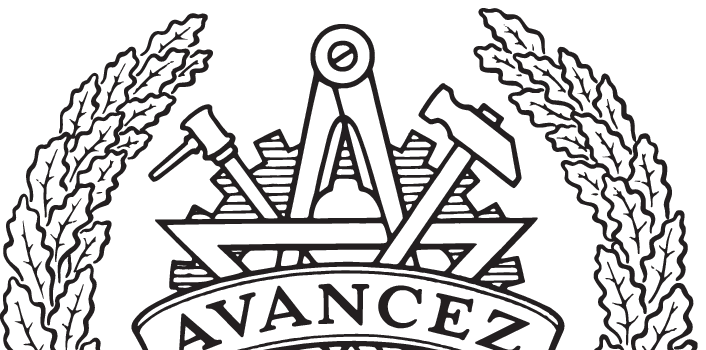}
\hfill $\begin{array}{r}\text{Gothenburg preprint} \\ \text{December,
    2012}\\$\quad$\end{array}$
\vskip-3mm
\noindent\makebox[\linewidth]{\rule{\textwidth}{0.4pt}}
\vspace{\stretch{1}}
\begin{center}
{\LARGE Loop amplitudes in maximal supergravity\\ \vspace{0.1cm} with manifest supersymmetry}\\
\vspace{\stretch{1}}
{\normalsize Martin Cederwall \\ Anna Karlsson}\\
\vspace{\stretch{1}}
{\small Fundamental Physics \\ Chalmers University of Technology \\ SE 412 96 Gothenburg, Sweden }\\
\vspace{\stretch{1}}
\end{center}

\begin{abstract}\noindent
We present a description for amplitude diagrams in maximal
supergravities obtained by dimensional reduction from $D=11$, derived
from a field theory point of view using the pure spinor formalism. The
advantages of this approach are the manifest supersymmetry present in
the formalism, and the limited number of interaction terms in the
action. Furthermore, we investigate the conditions set by this
description in order for amplitudes in maximal supergravity to be
finite in the ultraviolet limit. Typically, there is an upper
limit to the
dimension, set by the loop order, which for an arbitrary number of
loops is no larger than two. In four dimensions,
the non-renormalisation power of the
formalism fails for the 7-loop contribution to the 4-point amplitude,
all of which is in clear agreement with
previous work.
\end{abstract}
\vspace{\stretch{2}}
\vfill
\noindent\makebox[\linewidth]{\rule{\textwidth}{0.4pt}}
{\tiny \texttt{email: martin.cederwall@chalmers.se, karann@chalmers.se}}
\thispagestyle{empty}
\newpage
\tableofcontents
\newpage
\section{Introduction}
In maximal supergravity, one remaining question concerns the behaviour
of amplitude diagrams, especially as to the finiteness of the
theory. While maximally supersymmetric Yang--Mills theory (SYM)
\cite{BSS} has proven to be perturbatively finite in four dimensions
\cite{BLN,Mandelstam,HST},
a corresponding statement is not available
for supergravity --- it is still unknown
whether or not $\mathcal{N}=8$ supergravity \cite{deWitNicolai} (the
dimensional reduction of $D=11$ supergravity
\cite{ElevenSG,ElevenSSSG,CremmerFerrara})
is perturbatively finite, or where in perturbation theory the first
divergences may set in.
The general conditions on the formulation
  in order for it to be finite still have to be shown.
The prevalent opinion seems to be that supergravity on its own it
not a sensible quantum theory, and that the ultraviolet completion is
string theory or M-theory, although there are dissenting views (see
e.g. ref. \cite{Kallosh}).
It is often assumed that the 4-point amplitude has a divergence
  at 7 loops 
\cite{V,BjornssonGreen,JB,BEFKMS}.

The techniques for determining the properties above typically
  make use of maximally supersymmetric quantum field theories, which
  have been of considerable interest lately. However, the low energy
  limit of M-theory, maximal supergravity, appears to diverge in the
  ultraviolet limit for $D\geq2$
  \cite{BjornssonGreen,JB,BossardHoweStelle}, 
though explicit
  calculations for the four-graviton amplitude in four dimensions
  ($\mathcal{N}=8$) show no divergences up to four loops
\cite{GSB,BDDPR,BCDJKR,Z.B},  
which is as far as detailed
  examinations have been conducted, so far. This latter procedure is
  complicated and time consuming, which is why it is interesting to
  explore alternative descriptions. One such is to use pure spinors as
  in ref. \cite{JB}.

The pure spinor formalism has been of use in maximally
  supersymmetric theories and string theory for some time, its main advantage being the fact that it provides a way of
  having manifest supersymmetry present.
In a maximally supersymmetric model, the ordinary construction in
terms of superfields works differently than in less supersymmetric
situations, in that supersymmetry transformations close under
commutation only modulo equations of motion. This means that the
supermultiplets are representations of supersymmetry only on-shell, and
there is no way of formulating an action principle with these
superfields only, since they do not provide auxiliary fields for
off-shell supersymmetry. Pure spinor superfield theory provides a way to encode the
  traditional superspace constraints, leading to the on-shell multiplets, as
  equations of motion. This requires the introduction of new bosonic
  variables, the pure spinors.

It was first observed in ref. \cite{N}, that a certain representation
of the Lorentz group was needed in order to lift the equations of
motion of $D=10$ SYM, and that this representation could be associated
with pure spinors. Other early works \cite{H1,H2} noted a connection
between supersymmetry and pure spinors, both for gauge theory and
supergravity. The breakthrough came with the realisation by Berkovits
that the long
sought for covariant quantisation of manifestly supersymmetric superstring
(or superparticle) theory could be achieved using a set of variables
that indeed are pure spinors \cite{B1,B2}.
Independently of this advance, in connection to the search for
higher-derivative deformations of maximally supersymmetric models
\cite{CGNT,CGNN,CNT1,CNT2,HoweTsimpis},
it was realised that the on-shell property could be
understood in terms of cohomology\footnote{Originally termed
    ``spinorial cohomology'', it was soon realised that this is
    exactly what is obtained from a pure spinor formulation.} \cite{CNT3}.
The ensuing field--antifield structure of maximally supersymmetric field
theory has been investigated and actions have been given for virtually
all models that do not involve self-dual fields
\cite{B2,MS,CederwallBLG,CederwallABJM,C1,C2,CederwallReview},
including supergravity.

A generic feature of interactions in pure spinor field theory is
  that the maximal order of the interaction terms is lower than in a
  component formulation. In supergravity, this even makes the
  interactions polynomial \cite{C2}, a phenomenon which has also been
  observed for the abelian Born--Infeld theory \cite{CK}.
Having a full interacting field theory with a limited number of
interaction terms is a great advantage when it comes to perturbation
theory, which we will use in the present paper. It indeed gives a
more direct explanation of many of the properties of amplitudes
\cite{BDDPR,BCDJKR,J1,J2,KLT}
that are
difficult to discern from a component action.
Previous approaches to perturbation theory using pure spinors,
e.g. ref. \cite{JB}, have
relied on first quantisation, a construction of a
  description with pure spinors through comparisons with string
  theory. In such a formulation, the consistency of every vertex (and
  eventually the amplitudes) with respect to the symmetries of the
  theory must be checked, something which the
  consistency of the classical action takes care of in a field
  theory. We have been much inspired by the work of ref. \cite{JB},
  and indeed use similar techniques in the concrete calculations. The
  use of $D=10$ pure
  spinors for amplitude calculations
in refs.
\cite{BjornssonGreen,JB,Berkovits&Nekrasov,AB,GV},
with its separation of
left- and right-movers is probably the ideal realisation of the
Kawai--Lewellen--Tye relations \cite{KLT}, but does not allow for a
field theory formulation, in contrast to the $D=11$ pure spinors used
here, which only calls for a single BRST operator.

The paper is organised as follows. In order to apply the pure
  spinor formalism of $D=11$
supergravity to the construction
  of amplitude diagrams, we begin by going through the key concepts of
  the pure spinor formalism in relation to supergravity in section
  \ref{sec.Char}. These are generally well known, but crucial for the
  subsequent part, section 3, where we address the question of how to
  describe amplitude diagrams, both tree diagrams and loop
  diagrams. The main ingredients for
  the construction of amplitudes are either given by the action, or by quite straightforward generalisations to $D=11$ pure spinors of
  techniques known from $D=10$.

The last part of the amplitude description though, consists of the analysis of the detailed
properties of loop amplitudes. These are intricate, but of great use in section
\ref{sec.UV}, the part on the UV behaviour of the amplitude
diagrams. Despite the different formalism, our final results are in
  complete agreement with those of refs. \cite{BjornssonGreen,JB}. 
However, our analysis might
  not be exhaustive, and might give at hand more properties, not yet
  noted, despite the intricateness of the 
description, as is noted in section \ref{sec.Con}, where our results
are summarised and discussed. A few useful spinor identities are listed
in appendix \ref{app.Fierz}, the zero-mode cohomology of the pure
spinor superfield is given in appendix \ref{app.zmc} and appendix
\ref{app.Pro} contains a derivation necessary for the standard
procedure of general regularisation of amplitude diagrams to be
applicable to supergravity.

\section{Characteristics of the pure spinor formalism}\label{sec.Char}
The advantage of the pure spinor formalism is that it provides an
off-shell formulation\footnote{That is, a formulation which can
  describe fields that do not obey the equations of motion, and can
  provide an action principle.} for maximally supersymmetric
theories. Typically the fields depend not only on the ordinary
  superspace coordinates, but also on a bosonic spinor
$\la^\alpha$ which is pure in the sense that it obeys the constraint
$\la\g_a\la=0$.

The reason for the introduction of the pure spinor $\la^\alpha$ is
that the equations of motion for the physical fields of a free, maximally
supersymmetric theory tend to follow a pattern which is reproduced by
\begin{equation}
Q=\la^\alpha D_\alpha=\la D
\end{equation}
where $D_\alpha$ is the covariant fermionic derivative, acting on a pure spinor superfield $\psi=\la^{\alpha_1}\ldots\la^{\alpha_n}
C_{\alpha_1 \ldots \alpha_n}(x,\theta)$. $C$ here is a superfield of the
original theory, dependent only on the superspace coordinates, which
contains all the physical fields of the
theory. The parts of $Q\psi$ which vanish due to the pure spinor
constraint
correspond exactly to the terms which are absent in the original
equations of motion\footnote{From the point of view of a
    superspace gauge theory, this happens due to conventional
    constraints (see e.g. ref. \cite{DragonWindow} for a full
    discussion). These are exactly what is needed in order to describe a
    theory in terms of the lowest-dimensional superfield. In the
    present case $C_{\alpha\beta\gamma}$ is the part of the superspace
    3-form with only
    fermionic indices.}. The linearised
dynamics of the original (free) theory is captured in full by one
singe equation of motion of the free enlarged theory: $Q\psi=0$.

In $D=11$ supergravity, this construction appears as
$\psi=\la^\alpha\la^\beta\la^\g C_{\alpha\beta\g}(x,\theta)$
\cite{C1}. The difference from the original theory is the presence of
the pure spinor, which is endowed with ghost number 1 in order to
separate physical fields and variables from the original theory (with
ghost number zero) from other components of the enlarged theory.
To be precise, these contain the set of ghosts and antifields
  appropriate for the theory, with the correct ghost numbers.
The general field $\psi(x,\theta,\la)$ (the only field present)
is fermionic, with ghost number 3 and no free indices, and has as
its $\la$- and $\theta$-independent part the ghost for ghost for
ghost of the tensor
gauge symmetry. It includes, in
  addition to the unconstrained superfield
  $C_{\alpha\beta\gamma}$ of ghost number zero, also
  superfields containing the relevant ghosts and antifields of the
  theory (and more).

In short, $Q$ can be recognised as a nilpotent operator, and the
cohomology of $Q$ represents the free on-shell
fields, including ghosts and antifields. The actual field content
can be read off in table \ref{table.CPsi} in appendix
\ref{app.zmc}. There, the Lorentz representations of the zero-mode
cohomology of $Q$ is given. The proper cohomology of $Q$ contains
these fields, constrained by some differential equation (equation of
motion) which also is in the zero-mode cohomology. Through a
separation of terms of ghost number zero from the rest, the enlarged
theory gives at hand results in the original theory. At the same time,
the presence of $Q$ enables off-shell calculations. The details of
this --- the main concepts of the formalism --- is what we will begin by
describing, below, before dealing with the details of the formalism.

In the end, the full action of $D=11$ supergravity contains more
than the kinetic operator $Q$. In addition to the term of the free
theory, describing the free propagation of a field, it contains a
3-point and a 4-point coupling\footnote{Note that the overall
  constant of the action, here displayed as the coupling constant
  $\kappa^{-2}$, depends
  on the normalisation of $\psi$. For
canonically normalised fields, it will appear in the
interaction terms instead.}:
\cite{C2}
\begin{equation}\label{eq.action}
S=\frac{1}{\kappa^2}\int[\mathrm{d}Z]\big(\frac{1}{2}\psi Q\psi +
\frac{1}{6}(\la\g_{ab}\la)(1-\frac{3}{2}T\psi)\psi R^a\psi R^b\psi
\end{equation}
The precise meaning of the operators involved, as well as the
  integration measure, will be elucidated later in this section. Importantly, the pure spinor formalism transforms the supergravity
action into a polynomial expression. The quartic term is the last term
and moreover, the operator $T$ is nilpotent so that a 4-point vertex
at most can appear once in a diagram. However, though the pure spinor
formalism simplifies the interaction terms in such a way, some parts
remain complicated. For example, it is not known how to write down a
solution for $\psi$. Although the formulation enjoys the full gauge
symmetry of $D=11$ supergravity (diffeomorphisms, local symmetry
and tensor gauge symmetry), the geometrical picture is obscured, as is
the background invariance. In a way, the complications have only been
transferred between different parts of the theory. For perturbative
theory though, the action above is ideal.

The parts of this action is what will be described below, after the
main concepts have been presented. To begin with, we discuss the
formalism with respect to what variables there are, including a
discussion on what the pure spinor really is: the solution to the pure
spinor constraint. Notably, the action is expressed in the so called
non-minimal formalism, which is what is usually used for the pure
spinor formalism, since the introduction of two more sets of
variables (apart
from the pure spinor mentioned above) enables integration.
The introduction of the concept of integration, and
the ensuing clarification of the importance of the non-minimal
formalism, is what is described right after the non-minimal
superspace.

The final part of the formalism we present, before heading off for new
theoretical aspects concerning the construction of amplitude diagrams,
is the operators involved in the action. It should be noted that
(most) concepts in this section are well known, though the treatment
of the solution to the pure spinor constraint in this formalism, see
subsection \ref{subsub.sol}, only has been presented to linearised
order in $\Omega$ before \cite{AABN}. Here, it will be treated in
full.

\subsection{The heart of the formalism}
The identification of the $Q$ described above is performed in a very
ad hoc manner. However, the theory it gives rise to is very well
defined. As the covariant derivative $D_\alpha$ obeys
$\{D_\alpha,D_\beta\}=-2\g^a_{\alpha\beta}\partial_a$ in flat space
and similar relations for the superspace torsion and curvature,
annihilated by the pure spinor condition, hold
for other backgrounds\footnote{In this paper, the background will be
  assumed to be flat, since we are aiming at Minkowski space
  amplitudes. The main difference with another background present
  would be that the other operators of the theory  would need to be
  constructed from covariant derivatives while taking the background
  into consideration. Such a procedure would be a bit more intricate,
  though probably entirely feasible. \cite{C2}}, $Q$ can be recognised
as a nilpotent operator, a BRST operator. This operator is not a
supersymmetry transformation operator, but commutes with all flat
superspace Killing spinors so that the global supersymmetry
transformations close both on- and off-shell. Essentially, the action
is recognised to give at hand a theory with the correct properties, so
that it represents a working action for the free theory. It has to be
remembered, though, that ghosts and antifields are included, and need
to be treated properly by gauge fixing.

The BRST construction concerns free theories, but it has a natural
extension to a theory of interactions, a field theory formulation: the
Batalin-Vilkovisky (BV) formalism \cite{BV,FHM}. There, the BRST
symmetry is
central, but while the BRST formulation can be compared to a first
quantised theory, the BV represents a second quantised
version (a field theory), a change which usually requires the
introduction of new fields (ghosts and antifields). The alteration
corresponds to exchanging the BRST symmetry, where $Q$ gives rise to the
variations, for the BV symmetry, where the generalised action
gives rise to the variations of the fields:
$(S,\psi)=\delta\psi$.

This antibracket usually is expressed in the introduced
fields and antifields.
However, all those fields are already present in the field
$\psi$, which is its own antifield, so for the pure spinor
formalism, the extension to a BV formalism is performed through
another ad hoc recognition of how to imitate the BV formalism. The
only antibracket possible to write down, where the field $\psi$ is
self-conjugate with respect to the antibracket, is:
\begin{equation}\label{eq.master}
(A,B)\sim \int \frac{\delta A}{\delta \psi}\frac{\delta B}{\delta \psi} [\mathrm{d}Z]
\end{equation}
This simple construction turns out to be the correct one, which can be checked on the component fields appearing in the cohomology listed in appendix \ref{app.zmc}.

This concludes the recognition of what the pure spinor formalism
really is. The expression for the equation of motion for a field
$\psi$ is:
\begin{equation}\label{eq.EQMS}
(S,\psi)=0
\end{equation}
If a pure spinor superfield $\psi$ fulfils this equation, and thus
obeys the equation of motion, it is termed to be on-shell. Gauge
invariance for a field is represented by the exact same expression, as
is typical in a BV formalism. Furthermore, the master equation which
the action must obey is
\begin{equation}
(S,S)=0
\end{equation}
which also gives at hand the form
of the action once the 3-point coupling has been deduced from certain
properties of the theory, a process which will be described below.

Most importantly, the BV formalism makes it possible to perform off-shell
calculations. It incorporates the alterations that are necessary in
order to bring the original on-shell theory off-shell, where
interactions can be taken into account in a manifestly supersymmetric
formalism. The BV formalism incorporates far more than the BRST
version through the change of the symmetry transformations. There are
many off-shell field components that take part in the interactions, at
the same time as the free physical fields can be expressed in the
minimal formalism and interpreted with respect to the original
theory. Furthermore, gauge symmetry is manifest throughout the
process, just as for the SYM theory, so at the point when the gauge
needs to be fixed, no extra procedure is necessary in order to check
gauge invariance, which is always present. These are considerable
strengths of the formalism.

\subsection{Superspace in the non-minimal formalism}\label{subsec.formalism}
In $D=11$, the coordinates which describe superspace consist of
11 bosonic variables ($x^a$) and 32 fermionic spinors
($\theta_{\alpha}$). They are symplectic, so $(\la \theta)=-(\theta \la)$, as they are connected by
the asymmetric tensor\footnote{For more details, e.g. the general
  Fierz identity of  four components: $(AB)(CD)$, have a look at
  appendix \ref{app.Fierz}.} $\varepsilon^{\alpha\beta}$. Furthermore,
the dimension of a variable $x$ is usually set to $-1$, so that the
corresponding for a spinor coordinate is $-1/2$.
In addition, in the pure spinor
formalism there is the pure spinor
$\la_\alpha$,
which is bosonic and has 9
degrees of freedom less than ordinary spinors due to the pure spinor
condition, i.e., 23 independent ones\footnote{We stick to
    to the terminology ``pure spinor'' for the algebraically constrained
    object $\la$ with $\la\g^a\la=0$,
    although it clearly differs from the traditional
    mathematical definition, which only makes sense in an even number
    of dimensions.}. In the non-minimal formalism there are also two other
variables, introduced due to their properties: $\lb_\alpha$ and
$r_\alpha$. The first of these is bosonic with ghost number $-1$,
whereas the second is fermionic with ghost number zero, and they obey
$(\lb\g^a\lb)=(\lb\g^a r)=0$. Both are of dimension $1/2$, so that the
set $(r,\lb)$ has quantum numbers conjugate to those of
$(\theta,\la)$. Note that at the introduction of the non-minimal variables, the BRST operator has to be modified in order for the cohomology not to be
changed, see subsection \ref{subsub.Q}.

Before moving on to the finer points of the theory, a brief
explanation of notation is convenient as well. To begin with, the
derivative of $\la^\alpha$ is denoted by $\omega_{\alpha}$. The
corresponding for $\lb_\alpha$ is $\bar{\omega}^\alpha$, which
illustrates an analogy with barred and unbarred quantities which is
used throughout this paper. The derivative with respect to $r_\alpha$ on the
other hand, is denoted by $s^\alpha$. Moreover, all expressions must
be gauge invariant, which is why all these derivatives invariably show
up in the combinations of \cite{B3}:
\begin{equation}\label{eq.gaugeN}
\begin{array}{lll}
N=\la\omega& \bar{N}=\lb\bar{\omega}+rs&S=\lb s\\
N_{ab}=\la\g_{ab}\omega\quad&\bar{N}_{ab}=\lb\g_{ab}\bar{\omega}+r\g_{ab}s\quad&S_{ab}=\bar{\la}\g_{ab}s
\end{array}
\end{equation}
Further notation consists of the two scalars possible to form from the
pure spinors in the  non-minimal formalism \cite{C1}:
\begin{equation}\label{eq.scalars}
\xi=(\la\lb)\qquad \eta=(\la\g^{ab}\la)(\lb\g_{ab}\lb)
\end{equation}
The last is not to confuse with $\eta^{ab}$, which simply lowers or
raises indices.

\subsubsection{The solution to the pure spinor constraint}\label{subsub.sol}
To wrap up the properties of $D=11$ superspace and the pure
spinor constraint, we now turn to the solution of that same
constraint. The procedure for finding out that the pure spinor has 23
degrees of freedom goes as follows.

A spinor in $D=11$ decomposes as
${\bf16}\oplus{\bf\overline{16}}$ when $so(11)\rightarrow so(10)$, and
thus becomes
\begin{equation}
{\bf32}\rightarrow
{\bf1}_{-5/2}\oplus{\bf5}_{-3/2}\oplus{\bf10}_{-1/2}
\oplus{\bf\overline{10}}_{1/2}\oplus{\bf\overline5}_{3/2}
\oplus{\bf1}_{5/2}
\end{equation}
under the further decomposition $so(10)\rightarrow su(5)\oplus u(1)$. Such a spinor contains all forms:
\begin{equation}
\la=\bigoplus\limits_{p=0}^5\la_p
\end{equation}
Some of these will be determined entirely by the pure spinor
condition, a condition which however looks a bit different in this
notation. The 10-dimensional $\g$-matrices act as\footnote{For
  practical purposes, $\omega$ is a 1-form and $\iota_v$ represents
  the contraction with the vector field $v$, reducing the form degree of the form it acts on by one.} $\omega\w$ and $\iota_v$ and in addition to those, there is the $\g^{11}=(-1)^p$, $p$ being the form degree. In total, the $so(11)$-invariant symplectic spinor scalar product is:
\begin{equation}
<A,B>\,={\star}\sum\limits_{p=0}^5(-1)^{\frac{p(p+1)}{2}}A_p\w B_{5-p}
\end{equation}
With this, the pure spinor constraint translates into:
\begin{equation}\label{PureElevenEq}
<\la,(-1)^p\la>\,=\,<\la,\omega\w\la>\,=\,<\la,\iota_v\la>\,=0
\end{equation}
As $\la_0$ is just a scalar factor and the wedge product is symmetric for $\la_{\text{even}}$, whereas the opposite is true in between $\la_{\text{odd}}$'s, the first and second equation can be used to solve for $\la_5$ and $\la_4$ respectively, in terms of the other $\la_n$'s:
\begin{equation}\label{LambdaFourFive}
\begin{array}{l}
\la_4=\la_0^{-1}(-\la_1\w\la_3+\frac{1}{2}\la_2\w\la_2)\\
\la_5=\la_0^{-1}\la_2\w\la_3-\frac{1}{2}\la_0^{-2}\la_1\w\la_2\w\la_2
\end{array}
\end{equation}
The last condition on the other hand, the third of the equations in eq. \eqref{PureElevenEq}, corresponds to:
\begin{equation}
\begin{array}{r}
-2\iota_v\la_1\w\la_5+2\iota_v\la_2\w\la_4-\iota_v\la_3\w\la_3=0\\
\stackrel{\eqref{LambdaFourFive}}{\Longleftrightarrow}
\iota_v(\la_3-\la_0^{-1}\la_1\w\la_2)\w(\la_3-\la_0^{-1}\la_1\w\la_2)=0
\end{array}
\end{equation}
Here, we have rearranged the expression in order to make the solution
visibly clearer. Its solution contains a 3-form $\Omega$:
\begin{equation}\label{GrassmannianEq}
\la_3=\la_0^{-1}\la_1\w\la_2+\Omega\li, \quad \iota_v\Omega\w\Omega=0
\end{equation}
This means that the pure spinor is described in full by $\la_0$, $\la_1$, $\la_2$ and $\Omega$, where the condition on $\Omega$ needs to be examined further.

The solutions for $\Omega$ describe, modulo a scale, the Grassmannian $Gr(2,5)={SU(5)/ S(U(3)\times U(2))}$ of 2-planes in ${\mathbb C}^5$. The 16-parameter solution with $\Omega=0$ is the space ${SO(12)/ SU(6)}\times{\mathbb R}_+$ of 12-dimensional pure spinors, and the full solution is a bundle over that space with fibre $Gr(2,5)\times{\mathbb C}$. For a parametrisation of the solutions, we break $su(5)\rightarrow su(3)\oplus su(2)\oplus u(1)$, so that $\Omega$ splits into $\Omega_{(3,0)}$, $\Omega_{(2,1)}$ and $\Omega_{(1,2)}$, where the form degrees are with respect to the $su(3)$ and $su(2)$ parts, respectively. Their charges under the ``new'' $u(1)$ are $-1$, $-\frac{1}{6}$ and $\frac{2}{3}$, set by the assignment of the charge $\frac{1}{3}$ to an $su(3)$ vector and the charge $-\frac{1}{2}$ to an $su(2)$ vector. Furthermore, it turns out to be convenient to dualise the $su(3)$ form indices to vector indices:
\begin{equation}
{\star}_3\Omega=\omega+\Omega^1{}_1+\Omega^2{}_2
\end{equation}
In order to use this though, we need to relate wedge products of 1-forms with forms to contractions of 1-forms with the dual multi-vector, and correspondingly for contractions with vectors. For a general dimension $n$ and form degree $p$ we have
\begin{equation}
\begin{array}{r}
{\star}(\omega\w A_p)=(-1)^{n-p+1}\iota_\omega{\star}A_p\\
{\star}(\iota_v A)=(-1)^{n-p}v\w{\star}A_p
\end{array}
\end{equation}
where wedge products between vectors are defined just as between forms. We also need the general relation between scalars:
\begin{equation}
{\star}(A_p\w B_{n-p})={\star}({\star}A_p\w{\star}B_{n-p})
\end{equation}

Now, the constraint in eq. \eqref{GrassmannianEq} splits into two sets of equations, since the vector $v$ decomposes in $v^{(1,0)}$ and $v^{(0,1)}$. The one containing $v^{(1,0)}$ is:
\begin{equation}\label{XiTwoTwo}
2\omega\Omega^2{}_2-\Omega^1{}_1\w\Omega^1{}_1=0 \quad\Rightarrow\quad \Omega^2{}_2=\frac{1}{2}\omega^{-1}\Omega^1{}_1\w\Omega^1{}_1
\end{equation}
The solution
happens to be just what it takes for the equation containing
$v^{(0,1)}$ to be automatically satisfied. In total, this gives at
hand that the complete solution of the $D=11$ pure spinor constraint
is parameterised by the 23 variables
$(\la_0,\la_1,\la_2;\omega,\Omega^1{}_1)$.
\\ \\
We now turn to the expression for the volume form for the pure
spinor. It is expected to be given by $[\mathrm{d}\la]=\la_0^\alpha
\id\la_0\id^5\la_1\id^{10}\la_2\li\omega^\beta \id\omega\id^6{\Omega^1}_1$ for some numbers $\alpha$ and $\beta$, but a counting
of the two $u(1)$ charges immediately gives at hand that $\alpha=-5$
and $\beta=-2$. What remains is to show the $so(11)$ invariance of the
volume form. The first step in this is to show the $su(5)$ invariance
of $\omega^{-2} d\omega d^6\Omega^1{}_1$, after which one can proceed
to the full volume form. The calculations are a bit tedious, but everything
works out. The volume form is thus:
\begin{equation}
[\mathrm{d}\la]=\la_0^{-5} \id\la_0\id^5\la_1\id^{10}\la_2\li\omega^{-2} \id\omega \id^6\Omega^1{}_1
\label{eq.topform}
\end{equation}
Here, the seven inverse powers of the pure spinor reflects the highest
(ghost antifield) cohomology at $\la^7$. The two powers of $\omega$
agrees with the concrete form of the ghost antifield cohomology in
refs. \cite{C1,C2}, carrying exactly two factors of
$(\la\g^{(2)}\la)$.

As a representative pure spinor $\la$ in the generic, 22-dimensional orbit under $Spin(11)$, i.e., one with
$(\la\g^{ab}\la)\neq0$, one can take the pure spinor with only the two
numbers $\la_0$ and $\omega$ non-vanishing, and
$\la_1=\la_2=\Omega_{(2,1)}=0$, implying
$\Omega_{(1,2)}=\la_4=\la_5=0$. For this representative,
$(\la\g_{ab}\la)$ only has a single non-vanishing component, when the
indices are upper $su(2)$ indices, and
$(\la\g^{a'b'}\la)\sim\la_0\omega\e^{a'b'}$. The correspondence with the scalar quantities defined in eq. \eqref{eq.scalars} is $\xi\sim|\la_0|^2$, $\eta\sim|\la_0|^2|\omega|^2$. A representative with $\omega=0$ is in the 15-dimensional orbit of
$D=12$ pure spinors.

The space of pure spinors can be equipped with a Calabi--Yau
structure, where the holomorphic top form is precisely the one given
in eq. (\ref{eq.topform}). This is in analogy with the ten-dimensional
case \cite{CederwallGeometry}, where more details about the metric and
K\"ahler structures were given. The integration described in the
following subsection will have a natural interpretation in terms of
the Calabi--Yau geometry on pure spinor space.

\subsubsection{Integration and divergences}\label{sub.IntADiv}
As stated above, the volume form for the pure spinor contains seven inverse powers of the pure spinor, so that the combined volume form for the superspace variables is:
\begin{equation}
[\mathrm{d}z]=\la^{-7}\mathrm{d}^{11} x\li \mathrm{d}^{32} \theta \li\mathrm{d}^{23} \la
\end{equation}
However, this does not suffice as a measure factor for the action, as
it has the wrong ghost number and dimensional properties. A proper
$[\mathrm{d}Z]$ in eq. \eqref{eq.action} must have ghost number $-7$
and dimension $-3$, whereas $[\mathrm{d}z]$ has ghost number 16 and
dimension $-3$. Furthermore, the the addition of a function of the
superspace variables to $[\mathrm{d}z]$ would not do, since such a
measure would be degenerate. It would exclude a lot of states in
$\psi$, due to $\psi$ being an expansion in the variables, some parts
of which would have higher powers of the superspace variables than
that type of measure could take. Neither is it a priori clear how
to perform integration over the 23 holomorphic variables $\la$. A treatment using only these variables typically has to be either
non-manifestly covariant or involve picture changing operators
\cite{BerkovitsMinimalLoop,BedoyaGomez}.

The trick is instead to introduce the non-minimal variables \cite{B3}
($\lb_\alpha$, $r_\alpha$) and extend the formalism into the
non-minimal formalism, which is a possibility due to the properties of
$Q$, which will be discussed below. The previously mentioned
conditions on the non-minimal variables: $\lb_\alpha$ a pure spinor
with ghost number $-1$ and $r_\alpha$ a fermionic spinor of ghost
number $0$ with $(\lb\g^ar)=0$, both of the same dimension ($1/2$), is just
what it takes in order to form a proper measure factor for the
action. In fact, $[\mathrm{d}\lb][\mathrm{d}r]\sim
\mathrm{d}^{23}\lb\li\mathrm{d}^{23}r$ since \cite{C1,BerkovitsMembrane,AGV}:
\begin{align}
[\mathrm{d}\la]\la^{\alpha_1}\ldots\la^{\alpha_7}=&\star\bar{T}^{\alpha_1\ldots\alpha_7}{}_{\beta_1\ldots\beta_{23}}\mathrm{d}\la^{\beta_1}\ldots
\mathrm{d}\la^{\beta_{23}}\nonumber\\
[\mathrm{d}\lb]\lb_{\alpha_1}\ldots\lb_{\alpha_7}=&\star
T_{\alpha_1\ldots\alpha_7}{}^{\beta_1\ldots\beta_{23}}\mathrm{d}\lb_{\beta_1}\ldots
\mathrm{d}\lb_{\beta_{23}}\label{eq.TMeasure}\\
[\mathrm{d}r]=\lb_{\alpha_1}\ldots\lb_{\alpha_7}&\star\bar{T}^{\alpha_1\ldots\alpha_7}{}_{\beta_1\ldots\beta_{23}}
{\frac{\partial}{\partial r_{\beta_1}}}\ldots{\frac{\partial}{\partial
    r_{\beta_{23}}}} \nonumber
\end{align}
The right properties for $[\mathrm{d}Z]$ are thus given by:
\begin{equation}
[\mathrm{d}Z]=\mathrm{d}^{11} x\li \mathrm{d}^{32} \theta
[\mathrm{d}\la][\mathrm{d}\lb][\mathrm{d}r]
\end{equation}
The precise tensor structure of this measure is implied by
the statement that the tensor\footnote{The $\star$ in
eq. \eqref{eq.TMeasure} implies dualisation from 9 to 23 antisymmetric
spinor indices.} $T$ is obtained as the Clebsch--Gordan coefficients for the
formation of a singlet from 7 symmetrised indices and 9
antisymmetrised, both groups of indices being in the irreducible module (020003)
\cite{C1}:
\begin{align}
&T_{\alpha_1\ldots\alpha_7}{}^{\beta_1\ldots\beta_9}\la^{\alpha_1}\ldots\la^{\alpha_7}
\theta_{\beta_1}\ldots\theta_{\beta_9}\nonumber\\
&\sim(\la\g^{ab}\la)(\la\g^{cd}\la)(\Lambda^{ijklm}\g^n\theta)(\theta\g_{abp}\theta)(\theta\g_{cdp}\theta)(\theta\g_{ijm}\theta)(\theta\g_{kln}\theta)
\label{eq.explT}
\end{align}
Here, $\Lambda^{ijklm}_\alpha$ is in (00003):
$\Lambda_\alpha^{ijklm}=\la_\alpha(\la\g^{ijklm}\la)-2(\g^{[ijk}\la)_\alpha(\la\g^{lm]}\la)$.
Alternatively, we have \cite{BerkovitsMembrane,AGV}:
\begin{align}
&T_{\alpha_1\ldots\alpha_7}{}^{\beta_1\ldots\beta_9}\la^{\alpha_1}\ldots\la^{\alpha_7}
\theta_{\beta_1}\ldots\theta_{\beta_9}\nonumber\\
&\sim\epsilon_{a_1\ldots
  a_{11}}(\la\g^{a_1}\theta)\ldots(\la\g^{a_7}\theta)
(\theta\g^{a_8a_9a_{10}a_{11}}\theta)
\end{align}

Moreover, it is worth mentioning that this full integration over non-minimal
pure spinor space has a very natural and elegant geometric
formulation \cite{Berkovits&Nekrasov,CederwallGeometry}. In the
``Dolbeault picture'', where the variable $r$ is identified with the
antiholomorphic one-form $d\lb$, any function of $\la$, $\lb$ and $r$
becomes a cochain with anti-holomorphic form indices. The integration
then is defined using the holomorphic top form $\Omega$ of eq.
(\ref{eq.topform}), and
is given as:
\begin{equation}
\int[\mathrm{d}Z]\psi=\int\Omega\wedge\psi
\end{equation}

Importantly, the measure factor has the property that it extracts
$\la^7$ for the
integration over the pure spinors, part of which is two factors
of $\la\g^{(2)}\la$ due to the form of $\star\bar{T}$
given in eq. \eqref{eq.explT}
\cite{C1}.
Furthermore, for a non-zero result the fermionic integrals
must be saturated, which means that the integrand must contribute with
$32$ $\theta$'s and $23$ $r$'s.

The latter represents a problem, as there are demands on the integrand
in order for the integral not to diverge as well. In particular, there
ought not be too many negative factors of the scalars in
eq. \eqref{eq.scalars}, which would represent a divergence at the
origin with respect to $(\la,\lb)$ and/or at the submanifold of complex
codimension 7 where $\eta$ vanishes. It is however convenient to
separate those two behaviours as well as to treat factors of
$\la\g^{(2)}\la$, which is why the latter divergence is described with
respect to $\sigma$: $\sigma^2\sim\eta/\xi^2$. Each factor of
$(\la\g^{ab}\la)$ and $(\bar{\la}\g^{ab}\bar{\la})$ goes like
$\sigma$, and since the submanifold has real codimension 14 and
  the measure removes two powers of $\sigma$, we arrive at
$[\mathrm{d}Z]\sim \sigma^{11}\mathrm{d}\sigma$. In total, the
volume form of the pure spinor thus sets the following conditions on
the components of an integrand in order for it to be finite: \cite{C1}
\begin{align}
\left\{\begin{array}{rl}
\xi^x\li, & x>-23 \\
\sigma^y\li\, & y>-12\label{eq.reqSG}
\end{array}\right.
\end{align}
Here, the first condition applies to an integrand $\la^7\xi^x$.
The requirements are a bit worrisome, since the operators of section
\ref{sub.Op} typically 
contain negative powers of $\eta$. When there is a lot of them
present, the integral may give as nonsensical a result as
$0\times\infty$.
typically diverging the most severely with respect to
$\sigma$.
On top of that, the part giving at hand a divergence can
just as well come from a too high power of $(\la,\lb)$ as well, in
the limit where $\xi$ approach infinity. Typically this is a sign of
that some part of the theory is missing, that is, the
regularisations for these types of divergences.

However, those properties are here considered to be more tightly
connected to the amplitude diagrams than the formalism concerning the
action, and will therefore not be presented until the next
section. Before any of that though, the operators of the theory need
to be presented, as the important building blocks they are.

\subsection{The operators of the action}\label{sub.Op}
At the heart of the formalism, as mentioned above, is the BRST
operator $Q$ which is identified in an ad hoc manner, yet gives at
hand a clearly defined and elegant BV formalism. It shapes the rest of
the operators, partly in what ways gauge fixing might be done, which
gives at hand a condition for the propagator, and most definitely in the
expressions for the vertices, which are described by the vertex
operators. Most of them turn out to be singular at $\eta=0$.

\subsubsection{The kinetic operator $Q$ and regulators}\label{subsub.Q}
The BRST operator $Q$, mentioned in the beginning of this section,
changes by an addition of the operator $\bar{\partial}$ with the
enlargement of the theory into the non-minimal formalism:
\begin{equation}
Q=\la D + r \bar{\omega} \equiv q+\bar{\partial}
\end{equation}
The notation $\bar\partial$ is due to the geometric identification of
the variable $r$ with the anti-holomorphic differential $d\lb$ (the
``Dolbeault picture'') \cite{Berkovits&Nekrasov}.
This modified $Q$ has the same cohomology as the initial $Q$, in the minimal
formalism, and the proper ghost number and dimensional properties,
which is why the alteration is
correct. Formally, the first statement is that the operator above has
a cohomology which in each class has a representative independent of
the non-minimal variables \cite{B3,HT}, so it incorporates the
non-minimal formalism while conserving the desired properties of the
minimal formalism. The properties of the original theory can still be
extracted.

$Q$ is the kinetic operator of the formalism, and there is one very
useful twist to it. The free cohomology of an expression\footnote{The
  free cohomology of $X$ is the parts of $X$ that obey $[Q,X\}=0$,
  with the equivalence relation $X\approx X+[Q,Y\}$,
  where the brackets are set by the statistics of $X$ and $Y$.
Compare with eq. \eqref{eq.EQMS}, which describes the cohomology of $\psi$.}
is unchanged at the introduction of a term which is $Q$-exact,
provided that all amplitude calculations are performed between
on-shell external states. This is generally assumed, which is why
there is a freedom of the formalism: $1\rightarrow
1+\{Q,\chi\}$. Typically, this shows up in the use of so called
regulators, which can be added or removed at any time:
\begin{equation}\label{eq.regulatorChi}
e^{\{Q,\chi\}}=1+\{Q,\chi\}+\ldots
\end{equation}
The only conditions on $\chi$ are such that the exponent makes sense,
e.g. with respect to dimension and ghost number, which is why $\chi$
must have ghost number $-1$. It may even introduce new variables (and
corresponding integration measures), in a way similar to the
introduction of the non-minimal variables. Consequently, $Q$ would
change correspondingly as well. As the terminology hints at,
regulators will be of great importance in
the regularisations to be described below. Due to the
$Q$-invariance of the theory, one can choose to treat the first
  term in the expansion of eq. \eqref{eq.regulatorChi} or the entire
  series in order to find the properties of the expression which the
regulator is part of. A third possibility is to use the first term
with the addition of the term in the series which allows for
constructive interpretation of the results. This will come in handy later on.

\subsubsection{The propagator and the $b$-ghost}
A typical feature of pure spinor field theories is that gauge fixing
cannot be performed in a way similar to what is done in classical
field theories. In a BV formalism, this typically requires the
identification of separate fields and antifields, some of which need
to be introduced in the non-minimal formalism. The process of gauge
fixing involves the
  elimination of the antifields, which with the use of a guage fixing
  fermion become expressed in terms of the fields.
The superfield $\psi$,
on the other hand, already contains all such fields\footnote{Recall
  that $\psi$ is its own antifield, as is implicit in
  eq. \eqref{eq.master}.}, including the ones necessary in order to
fix the gauge, but it is not possible to separate those fields from
the rest, without separating $\psi$ into components of definite
  ghost number. Instead, the principle is to imitate  string theory in a
covariant gauge known as the Siegel gauge \cite{Siegel}, as for the scalar
particle. This gives at hand a free propagator $b/p^2$:
\begin{equation}\label{eq.Qb}
\{Q,b\}=\partial^2 \qquad b\psi_{\text{on-shell}}=0
\end{equation}
Here, the equation to the right represents the choice of gauge, and
the one to the left is the correct behaviour of the $b$-ghost, the solution of
which by default ought to give at hand the property
$\{b,b\}=0$.

Now, the $b$-ghost is not present as a fundamental variable
in the theory to begin with since the BRST operator
was not constructed from a symmetry point of
view. A free field theory is ordinarily based on, or
related to, a first-quantised particle with a world-line
reparametrisation invariance. The constraint $p^2=0$ (or, in the
string theory case, the Virasoro constraint) then is the
generator of these reparametrisations,
and in effect, there is a corresponding $bc$ reparametrisation ghost
  system. In contrast, in maximally supersymmetric
theories, the equations of motion follow from the condition $Q\psi=0$,
and do not have to be imposed separately. In the absence of
reparametrisations, there is no $b$-ghost.
However, the first condition in eq. \eqref{eq.Qb} makes it
possible to construct a composite
$b$ operator with the correct properties. This is not surprising,
given the knowledge that the free fields are massless, and there
should exist a gauge choice where $p^2=0$ for the linearised
theory. The
construction is known from
maximal SYM \cite{AB} and superstring theory \cite{AB,B3}.
It is, in short, the $b$ operator which appears in
the propagator in the Siegel gauge.

The solution to $b$ is found from the ansatz $\{\la
D,b_0\}=\nolinebreak\partial^2$ and through an iteration of
$\{r\bar{\omega},b_{n}\}+\{\la D,b_{n+1}\}=0$, where the lowered
indices denote the numbers of $r$ involved in each part of the $b$
operator. It is most easily expressed using terms that are not manifestly
gauge invariant, i.e., that contain derivatives $\omega$ not in
  the combinations \eqref{eq.gaugeN},
although it can be expressed entirely in
the gauge invariant terms of eq. \eqref{eq.gaugeN} as well, which
means that it is a well defined part of the manifestly supersymmetric
formalism. Note 
though, that such terms may seem to be
$\bar{\partial}$-exact despite them not being so. A part
$[\bar{\partial},\rho]$ with
$\delta^a\rho=[(\la\g^a\la),\rho]\neq0$,
for example the term
$b_3$ in the $b$ operator shown below, cannot be shifted
away through the use of $Q$-exact terms, as $\rho$ is not a well
defined operator\footnote{This is a statement
    that deserves some moderation. As noted by several authors
    \cite{AABN,CederwallOperators}, there is
    a ``gauge invariant derivative'', $\tilde\omega_\alpha$, with
    $(\la\tilde\omega)=(\la\omega)$ and
    $(\la\g^{ab}\tilde\omega)=(\la\g^{ab}\omega)$. Replacing $\omega$
    by $\tilde\omega$ in $b_3$ clearly leaves it unchanged, but allows
  for it to be shifted away as a $\bar\partial$-exact term, since it is
  then gauge invariant without the
  leftmost factor $(\la\g^{ab}\la)$. This procedure can be continued
  until one indeed is left with a $b$ operator constructed entirely
  out of the minimal variables, a procedure which however only has
  been performed explicitly for the  $b$ operator in
  $D=10$. The singular behaviour is then replaced
by singularities of the type $(N+a)^{-1}$ for some positive integers
$a$. Such a minimal $b$ operator turns out to contain a term with
three fermionic derivatives, which is necessary for the correct
fermion 4-point function \cite{NathanPrivate}.
Although the construction is interesting, we have not chosen to use
it for further calculations.}.
\begin{align}\label{eq.b}
b&=\frac{1}{2}\eta^{-1}(\bar{\la}\g_{ab}\bar{\la})(\la\g^{ab}\g^iD)\partial_i\nonumber\\
&\quad-\frac{1}{12}\eta^{-2}L^{(1)}_{ab,cd}\Big[(\la\g^{abcij}\la)\eta^{dk}+\frac{2}{7}(\la\g^{acijk}\la)\eta^{bd}\Big]\Big[(D\g_{ijk}D)-24N_{[ij}\partial_{k]}\Big]\nonumber\\
&\quad+\frac{1}{2}\eta^{-3}L^{(2)}_{ab,cd,ef}\Big[(\la\g^{ab}\la)(\la{\g^{cde}}_{ij}\la)(\omega\g^{fij}D)\nonumber\\
&\quad\qquad\qquad+\eta^{df}(\la\g^{ab}\la)\Big(\frac{2}{7}(\la{\g^{ce}}_{ijk}\la)(\omega\g^{ijk}D)+\frac{9}{7}(\la{\g^{c}}_{i}\la)(\omega\g^{ei}D)\Big)\nonumber\\
&\quad\qquad\qquad+\frac{1}{3}(\la\g^{ce}\la)\Big((\la{\g^{fa}}_{ijk}\la)(\omega\g^{dbijk}D)-9\eta^{df}(\la{\g^{a}}_{i}\la)(\omega\g^{bi}D)\Big)\Big]\nonumber\\
&\quad+\frac{2}{3}\eta^{-4}L^{(3)}_{ab,cd,ef,gh}(\la\g^{ab}\la)(\la\g^{ce}\la)\Big[(\la{\g^{dg}}_{ijk}\la)(\omega\g^{fhijk}\omega)\nonumber\\
&\qquad\qquad\qquad\qquad\qquad\qquad\qquad\quad-9\eta^{df}(\la{\g^{g}}_{i}\la)(\omega\g^{hi}\omega)\Big]
\end{align}
This operator was essentially constructed in ref. \cite{UnP}.
For convenience, we have introduced a special short notation used
for the dependence on the non-minimal variables,
where $\lbracket\ldots\rbracket$ denotes an
antisymmetrisation between the $p+1$ antisymmetric pairs of indices:
\begin{equation}\label{eq.L}
L^{(p)}_{a_0b_0,a_1b_1,\ldots,a_pb_p}=(\lb\g_{\lbracket a_0b_0}\lb)
(\lb\g_{a_1b_1}r)\ldots(\lb\g_{a_pb_p\rbracket}r)
\end{equation}
One important feature of these tensors is that they have the following property:
\begin{equation}
[\bar{\partial},\eta^{-(p+1)}L^{(p)}_{a_0b_0,\ldots,a_pb_p}\li\}
=2(p+2)\eta^{-(p+2)}L^{(p+1)}_{ab,a_0b_0,\ldots,a_pb_p}\li(\la\g^{ab}\la)
\end{equation}

Note that the $b$ operator, as well as the operators in the action,
described in the following subsection, contains negative powers of
$\eta$ only. This generic property is not a priori obvious, and one
can indeed find an alternative expression for the $r$-independent part
of $b$:
\begin{equation}
b'_0=\frac12\xi^{-1}(\lb\g^aD)\partial_a
\end{equation}
This expression turns out to be $q$-equivalent to $b_0$, but it does
not, for some reason, allow for the analogous construction of a full
$b$ operator \cite{UnP}.
In ref. \cite{B3}, where the non-minimal variables were first
introduced, the topological aspects of the dependence on the pure
spinor variables were stressed. There, one can interpret the
topological property as the possibility to localise wave functions
near the origin of the space of $D=10$ pure spinors. Here, in
contrast, the localisation seems to take place at the singular locus
$\eta=0$, which is the space of twelve-dimensional pure spinors. The deliberations of ref. \cite{MovshevEleven} seem to aim at
  utilising this phenomenon.

Moreover, in constrast to the defining relation \eqref{eq.Qb}, we have
not performed the full calculation in
  order to show that $b^2=0$. After having
  checked the vanishing of a number of terms, we have instead chosen
to rely on the observation that there is no scalar cohomology
represented by an operator $\sigma$ such that a relation
$b^2=\{Q,\sigma\}$ could hold for $\sigma\neq0$.

As is also discernible from eq. \eqref{eq.Qb}, the propagator cannot act on
fields that are on-shell. Rather, it acts on of-shell fields produced
from vertex operators and external fields. This brings us to the
formation of the vertices of the theory, which of course is crucial
for the amplitude diagrams.

\subsubsection{The vertex operators $R^a$ and $T$}\label{subsub.vert}
In $D=11$ supergravity, there is one more pure spinor field than
$\psi$: $\phi^a$, which starts with the diffeomorphism ghost and has
the physical meaning of superspace geometry. Its cohomology coincides
with the on-shell fields obtained by solving the linearised torsion Bianchi
identities (after imposing conventional superspace constraints)
\cite{ElevenSSSG,CremmerFerrara,CGNT,CGNN}.
However, the two fields
represent the same physical degrees of freedom, so that it is possible
to express $\phi^a$ as $\phi^a=R^a\psi$: \cite{C1}
\begin{align}
R^a&= \eta^{-1}(\lb\g^{ab}\lb)\partial_b
-\eta^{-2}(\lb\g^{ab}\lb)(\lb\g^{cd}r)(\la\g_{bcd}D)\nonumber\\
&\qquad-16\eta^{-3}(\lb\g^{a[b}\lb)(\lb\g^{cd}r)
           (\lb\g^{e]f}r)(\la\g_{fb}\la)(\la\g_{cde}\omega)\qquad\qquad\quad\\
&=\eta^{-1}(\lb\g^{ab}\lb)\partial_b
  -\eta^{-2}L_{(1)}^{ab,cd}(\la\g_{bcd}D)
  -6\eta^{-3}L_{(2)}^{ab,cd,ef}(\la\g_{ef}\la)(\la\g_{bcd}\omega)
\nonumber
\end{align}
As for the $b$ operator, we here have chosen a presentation which is not
  manifestly gauge invariant with respect to the pure spinor
  constraint. For an expression in terms of gauge
  invariant operators, see ref. \cite{C2}.

The operator $R^a$ obeys $[Q,R^a]=0$ modulo terms of the form
$(\la\g^a{\cal O})$, which allows for it to represent
cohomology due to the fact that $\phi^a$ has an extra gauge invariance
such that $\phi^a\approx\phi^a +(\la\g^a\rho)$ for an arbitrary
$\rho(Z)$ \cite{C1}. Such ``shift symmetries'' are on equal footing with
the pure spinor constraint and generic for non-scalar spinor
superfields, such as $\phi^a$. They were discussed in detail
in ref. \cite{CK}, and also occur for the fields of refs.
\cite{CederwallBLG,CederwallABJM,DragonWindow,CederwallNilssonSix}.

The physical meaning of $\phi^a$, superspace geometry, is the clue to
its role in the action, as a part of the vertices. There are several
properties which make
\begin{equation}
S_3\sim\int[\mathrm{d}Z](\la\g_{ab}\la)\psi\phi^a\phi^b
\end{equation}
a good candidate for a 3-point coupling, which ought to be present in the full
theory. To begin with, the quantum numbers match as $R^a$ has
dimension $-2$ and ghost number 2. Moreover, the term
$(\la\g_{ab}\la)$ is exactly what is necessary for the correct
properties of antisymmetry and shift symmetry to be
present \cite{C1,CederwallOperators}. Also note that the
  3-form potential of $D=11$ supergravity is contained in the
  cohomology in $\psi$,
  while $\phi^a$ only contains it through its field strength
  4-form. It was checked explicitly in ref. \cite{C2} that the
  Chern--Simons term $\int C\wedge H\wedge H$ is correctly produced
  from the present 3-point coupling, but even without this test,
  already the
  uniqueness of the deformation of the Pauli--Fierz action for
  linearised gravity \cite{Boulanger}
assures that a consistent non-trivial 3-point coupling will
  be that of $D=11$ supergravity.

The requirement for the addition of this 3-point coupling, the full
expression of which can be seen in eq. \eqref{eq.action}, to the free
action is that the final action obeys the master equation. For this to happen,
no more than one additional term is needed: the 4-point coupling with
the operator $T$. More specifically, the condition \makebox{$(S,S)=0$} as in
eq. \eqref{eq.master} gives at hand a 4-point coupling in the action
as in eq. \eqref{eq.action} with the operator $T$: \cite{C2}
\begin{equation}
T=8\eta^{-3}(\lb\g^{ab}\lb)(\lb r)(rr)N_{ab}
\end{equation}
To be precise, it appears in the calculation of $(S,S)$
through a commutator of two $R^a$ operators as $(\la\g_{ab}\la)[R^a,R^b]=3\{Q,T\}$.
The most important feature of $T$ is that it is nilpotent. We have
that the factor\footnote{Note that the factor
    $(\lb r)$ shows up quite
  frequently in other expressions as well, for example in the
  propagator. Many terms (but not all) will therefore combine to zero,
  especially with a 4-point vertex present.} $(\lb r)(\lb r)$ equals
to zero, so $T$ and thus any 4-point vertex cannot show up in an
amplitude diagram more than once.

It is worth emphasising that the polynomial action thus obtained
  is the full action for $D=11$ supergravity, and not some
  approximation.
Typically, pure spinor superfield interactions tend to be of lower
order than component ones. This has been observed in SYM \cite{B2},
and in fact in all maximally supersymmetric models
with action formulations
\cite{CederwallBLG,CederwallABJM,C1,C2}.
A similar simplification from a
non-polynomial component action to a polynomial pure spinor superfield
action was shown to occur in the case of maximally supersymmetric
Born--Infeld theory \cite{CK}. Non-polynomiality of a component action
will arise on elimination of auxiliary fields.
The action still enjoys the full gauge invariance encoded in the
  master equation, part of which constitutes diffeomorphism symmetry,
  local supersymmetry and tensor gauge symmetry.
It is satisfying to see how many amplitude properties,
otherwise obtained through more indirect methods, can be given a direct
field theory interpretation.
\\ \\
Thus having concluded our description of the action of $D=11$ super\-gravity presented in eq. \eqref{eq.action}, we
now turn to how to put the formalism to use in relation to amplitude
diagrams.

\section{Amplitude diagrams}
The remaining part of this paper concerns the construction and
examination of amplitude diagrams in supergravity, with a starting
point from field theory and expressed in the pure spinor
formalism. Effectively, what might be used for this construction is
determined by the presented theory. External fields must be
represented by on-shell pure spinor superfields $\psi$ with components
of ghost number 0, 3- and 4-point
vertices ought to be described in full by the action in
eq. \eqref{eq.action} and each propagator ought to contain
$b/p^2$. What other features might be needed must be covered by the
operators and variables either present in the theory or possible to
introduce into it, $\delta$-functions and integrations
$[\mathrm{d}Z]$. Perhaps the most subtle and powerful feature of the
formalism is that so called regulators, $Q$-exact terms, may be added
through the insertion of $e^{\{Q,\chi\}}$, as described in subsection
\ref{subsec.formalism}.

We will begin this section by performing a thorough identification of
how to describe tree parts of amplitude diagrams in terms of the
ingredients described above. Part of this incorporates the concept of
regularisation with respect to the divergences such integrands
may present, a procedure which is required in order for finite results
to always be obtainable. As in the previous section, it should be
noted that this last procedure is well known, though the projector
used for the general regularisation (in this formalism) has not
been presented before, to our knowledge.

Following that, we will introduce loops and the regularisation
which every loop must be subject to in order for general amplitude
diagrams to have the slightest possibility to be finite --- a
regularisation already known from SYM in a first
quantised version \cite{JB,Berkovits&Nekrasov}, which in refs.
\cite{BjornssonGreen,JB} was generalised to supergravity as well. The broader
question of finiteness is to be (partly) addressed in the next
section, but the general properties will be sorted out in this one.

The last subsection concerns features of the amplitude diagrams,
beginning with the ways loop momenta split between different loops and
why bubbles and triangles cannot show up in the amplitude diagrams, in
consistency with previous results. We end by some notes on how to deal
with components exterior to a loop structure\footnote{By a loop
  structure we here denote a loop or loops, sharing propagators or
  connected by such. Any part which is attached to \emph{one} loop but
  does not contain or constitute a part of any loop itself, is not
  considered as a part of the loop structure. Rather, we term it a
  tree part connected to the loop structure by an ``outer leg''. This
  invariably ends in one or several external legs.} in relation to the
very same loop structure, before we move on to the next section, which
concerns an actual examination of the loop amplitudes, with respect to
their UV behaviour.

\subsection{Building blocks for tree diagrams}
The tree components are mostly given by the form of the action in
eq. \eqref{eq.action}. Beginning with the vertices, they basically are
to be read straight off that same equation. By construction, they
cannot be anything else, and they fulfil all required properties. The
propagator on the other hand needs a slight change in order to fit in,
and then there really are no other parts of the tree diagrams but the
general regularisations, which will be described in the next
subsection. Our main concern is the ultimate ultraviolet
  behaviour of the last remaining momentum integrals, so we will not
  care about details concerning overall normalisation of amplitudes,
  neither combinatorial factors of diagrams nor absolute
  normalisations of integrals over pure spinor space.

\begin{figure}
\begin{picture}(110,60)(-20,-5)
\put(-5,40){$a)$}
\put(0,0){\line(1,1){25}} \put(25,50){\line(0,-1){25}}
\put(25,25){\line(1,-1){25}}
\put(18,23){\vector(-1,-1){15}}\put(0,20){$R^i$}
\put(27,18){\vector(1,-1){15}}\put(20,1){$R^j$}
\put(-2,-2){\circle{5}}
\put(25,52){\circle{5}}
\put(52,-2){\circle{5}}
\put(27,26){$(\la\g_{ij}\la)[\mathrm{d}Z]$}
%
\put(95,40){$b)$}
\put(110,0){\line(1,1){25}}
\put(135,25){\line(1,1){25}}\put(135,25){\line(-1,1){25}}
\put(135,25){\line(1,-1){25}}
\put(128,23){\vector(-1,-1){15}}\put(110,20){$R^i$}
\put(137,18){\vector(1,-1){15}}\put(130,1){$R^j$}
\put(132,32){\vector(-1,1){15}}\put(130,40){$T$}
\put(143,24){$(\la\g_{ij}\la)[\mathrm{d}Z]$}
\put(108,-2){\circle{5}}
\put(108,52){\circle{5}}
\put(162,-2){\circle{5}}
\put(162,52){\circle{5}}
%
\put(215,40){$c)$}
\put(220,25){\line(1,0){35}}
\put(232,32){$\frac{b\delta}{p^2}$}
\put(217,25){\circle{5}}
\put(258,25){\circle{5}}
\put(265,22){$=$}
\put(282,25){\line(1,0){60}}
\put(309,22){x}
\put(292,32){$\frac{b\delta}{p^2}$}
\put(321,32){$\frac{b\delta}{p^2}$}
\put(300,10){$Q[\mathrm{d}Z]$}
\put(279,25){\circle{5}}
\put(345,25){\circle{5}}
\end{picture}\caption{The three illustrations above show the building
  blocks of the tree diagrams. The 3-point (a) and 4-point (b)
  vertices are depicted with their components, some of which in
  supergravity act on the fields which are to be attached. They can be on-shell, representing external legs, which is not true for the
propagator (c). Note also that two propagators can be connected with
$Q$ since \protect{$\{Q,b\}=p^2$} and $\{b,b\}=0$, which gives at hand
one propagator as depicted in (c). It is implicit in the notation on
the RHS of (c) that the two propagators and the $Q$ are formed out of
three different sets of variables, brought together by
$[\mathrm{d}Z]$'s and $\delta$-functions.}\label{vertices}
\end{figure}
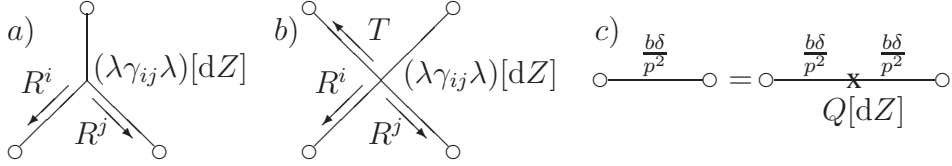

To begin with, the 3- and 4- point vertices connect
3 and 4 fields respectively, where each field is of ghost number 3. At
the same time, amplitude diagrams must be physical in the sense of
them having ghost number zero, as the action. The latter (and
dimensional properties) is what shapes the action and
the vertices so that the parts which are not fields $\psi$ have ghost
number $-9$ and $-12$ respectively. Part of both types of vertices are
two pure spinors and two $R^a$-operators, each of ghost number $-2$,
acting on the connected fields. There must also be an integration
$[\mathrm{d}Z]$, of ghost number $-7$, quite in analogy with ordinary
field theory where each vertex contains an integration over all of
space. This describes the 3-point vertex in full up to a number times
the coupling constant, which is given by the action and kept implicit
here. The 4-point vertex in addition contains $T$, acting on one of
connected fields just as the $R^a$'s. For an illustration of how this
works, see figure \ref{vertices}. What is important to remember is
that not two of the operators can act on the same field and give a
non-zero expression, but once that requirement is fulfilled, they can
be taken to act on any of the connected fields. The end result is the
same, a property which originates from partial integration in the
action (the operators $R^a$ and $T$ are first order in derivatives) and the fermionic statistics of the field $\psi$.

Moreover, a field needs to be outside of the cohomology in order to
propagate, so a propagator must be attached to two vertices. This
means that in order for all types of amplitude diagrams to end up with
ghost number zero in total, as should be, the total propagator which
connects different parts of amplitude diagrams must have ghost number
6. That way, each end of it can be considered to have ghost number 3,
as a field $\psi$ connected to a vertex would. This fits with the
total propagator consisting not only of $b$ (of ghost number $-1$) and
the factor $1/p^2$, but also a $\delta$-function, as depicted in
figure \ref{vertices}, since the $\delta$-function (with respect
  to the measure $[\mathrm{d}Z]$) must have ghost
number 7, as the ghost number of $[\mathrm{d}Z]$ is $-7$. However, the
presence of the $\delta$-function is in itself perfectly physical, as
the $b/p^2$ propagates the field coming from a vertex through
superspace, to another vertex. The connection is made by the
$\delta$-function, which makes the propagator ultralocal in the pure
spinor variables.

\begin{figure}
\begin{picture}(335,80)(-20,-15)
\put(-10,50){a)}
\put(0,0){\line(1,1){25}}
\put(25,50){\line(0,-1){25}}
\put(25,25){\line(1,-1){25}}
\put(27,26){$[\mathrm{d}Z]$}
\put(-10,6){$\psi_2$}
\put(12,45){$\psi_1$}
\put(50,6){$\psi_3$}
\put(75,50){b)}
\put(100,0){\line(1,1){25}}
\put(100,50){\line(1,-1){25}}
\put(155,25){\line(1,-1){25}}
\put(155,25){\line(1,1){25}}
\put(125,25){\line(1,0){30}}
\put(96,23){$[\mathrm{d}Z]$}
\put(90,6){$\psi_2$}
\put(90,41){$\psi_1$}
\put(135,32){$\frac{b\delta}{p^2}$}
\put(180,6){$\psi_4$}
\put(180,41){$\psi_3$}
\put(162,23){$[\mathrm{d}Z]$}
\put(208,50){c)}
\put(230,-15){\line(1,1){25}}
\put(230,65){\line(1,-1){25}}
\put(285,10){\line(1,-1){25}}
\put(285,40){\line(1,1){25}}
\put(255,40){\line(1,0){30}}
\put(255,10){\line(1,0){30}}
\put(255,10){\line(0,1){30}}
\put(285,10){\line(0,1){30}}
\put(229,36){$[\mathrm{d}Z]$}
\put(229,8){$[\mathrm{d}Z]$}
\put(220,-9){$\psi_2$}
\put(220,56){$\psi_1$}
\put(266,42){B}
\put(266,0){B}
\put(287,21){B}
\put(245,21){B}
\put(310,-9){$\psi_4$}
\put(310,56){$\psi_3$}
\put(290,36){$[\mathrm{d}Z]$}
\put(290,8){$[\mathrm{d}Z]$}
\end{picture}\caption{Some of the most basic tree and loop diagrams are the 3-point vertex (a) and the 4-point interaction amplitudes with none (b) and one loop (c) respectively, where the vertex operators have been left implicit. This shows the basic principles of connecting the constituents depicted in figure \ref{vertices} to each other and to external fields. In diagram (c), B denotes the total propagator $b\delta/p^2$. More usually though, the factors of $1/p^2$ are kept implicit, as well as integration measures and $\delta$-functions, which after all are integrated out before long.\label{simpAmplDiags}}
\end{figure}
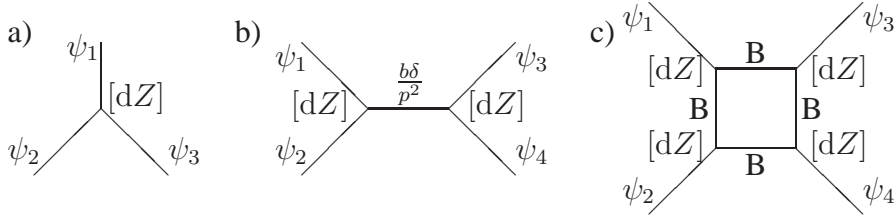

The components presented above suffice for the construction of tree
diagrams, as illustrated in figure \ref{simpAmplDiags}. At the
introduction of loops, a special regularisation is needed, which will
be treated in subsection \ref{subsec.loopReg}. Note though, that in
order to be finite, amplitude diagrams generally also need to be
regularised with respect to the pure spinors, as will be presented in
the next subsection.

\subsection{General regularisations}\label{subsec.genTransf}
At an integration over pure spinor space,
a general integrand typically
threatens to diverge in two different ways due to the presence of the
pure spinors $\la$ and $\bar{\la}$, as described in subsection
\ref{sub.IntADiv}. The first arises from $\xi\rightarrow\infty$. The
second contains the two different ways in which $\eta$ may approach
zero (at the origin or at the singular submanifold of
twelve-dimensional pure spinors),
so that negative powers of $\xi$ or $\eta$
diverge. The operators previously introduced typically contain
negative powers of $\eta$, which means that the first divergence to
occur is at the singular submanifold. This is the main cause
of divergent integrals.

Both types of divergences and
regularisations are known from maximal SYM \cite{JB,Berkovits&Nekrasov}, and
the procedure differs very little, despite the differences in the
  the singular submanifolds. The first type of divergence is
  regularised with the
introduction of a regulator:
\begin{equation}\label{eq.Genreg1}
e^{\{Q,\chi\}}=\Big[\chi=-\bar{\la}\theta\Big]=e^{-\la\bar{\la}-r\theta}
\end{equation}
This not only ensures a good convergence in the limit of infinity for
the bosonic spinors, but also saturates the fermionic integrals of
$[\mathrm{d}\theta][\mathrm{d}r]$. The regularisation, or a
  similar one, should be used already in the classical action in order
to produce sensible results for component fields. The bosonic part of
the regulator has a simple and attractive interpretation, where fields
are considered as sections of line bundles over the K\"ahler manifold
of pure spinors. This line bundle, a so called prequantum line bundle
(see e.g. ref. \cite{SpradlinVolovich}),
is equipped with a metric related to
the K\"ahler potential, which here is chosen to be $(\la\lb)$, the metric
obtained from embedding the pure spinor space in a flat $32_{\mathbb
  C}$-dimensional spinor space\footnote{This
    is not the K\"ahler potential
    corresponding to the Calabi--Yau metric \cite{CederwallGeometry},
but that is unimportant
  since the deformation is $Q$-exact. It is tempting to imagine that
  this connection to geometric
  quantisation could be continued, and that a non-commutativity on
  pure spinor space could provide natural regularisation also of other
potential divergences. We have not managed to show that this happens.}.

The latter divergence is superficial as well, provided that the
operators involved are well behaved enough. To begin with, note the conditions on an
integrand in order for it to be finite, listed in
eq. \eqref{eq.reqSG}. If an operator obeys those conditions, it can be
shown that it is BRST equivalent with an operator which is not
singular with respect to $(\xi,\eta)$, as in ref.
\cite{Berkovits&Nekrasov}
for SYM. Moreover, since we only are interested in
BRST-equivalent
classes of operators (the key feature of the introduction
of regulators in subsection \ref{subsub.Q}), the singular operator can
be exchanged for its regular BRST-equivalent.

The principle is that an integration over $[\mathrm{d}Z]$ (without
$[\mathrm{d}x]$) can remove the negative effective powers of $\xi$ and
$\sigma$ in the operator, so a new set of variables (with the same
properties as the first) is introduced and integrated over. The result
is an operator dependent on the non-minimal variables, which does not
contribute to any divergence at the origin or at the singular
submanifold. As the procedure can be performed for any number of
operators, and all operators fulfil eq. \eqref{eq.reqSG}, each and
every integrand has a representative of its orbit under BRST
equivalence which satisfies eq. \eqref{eq.reqSG}. That is why this
procedure, described below, regularises the divergence with
respect to negative powers of $(\xi,\eta)$.

A sketch of the procedure is as follows. With the introduction of two constant pure spinors $f_\alpha$ and $\bar{f_\alpha}$ (corresponding to a new set of $\la_\alpha$ and $\lb_\alpha$) a change of the operator might be performed: \cite{Berkovits&Nekrasov}
\begin{align}\label{eq.GuessOp}
O(\la,\bar{\la}) \leadsto &\int[\mathrm{d}f][\mathrm{d}\bar{f}]e^{-f\bar{f}}e^{i\varepsilon(fW+\bar{f}\bar{W})}O(\la,\bar{\la})=\nonumber\\
=&\Big[\la'=e^{i\varepsilon fW}\la\li, \li\bar{\la}'=e^{i\varepsilon \bar{f}\bar{W}}\bar{\la} \Big]=\\
=&\int[\mathrm{d}f][\mathrm{d}\bar{f}]e^{-f\bar{f}}O(\la',\bar{\la}')\nonumber
\end{align}
Here, ($\la'$, $\bar{\la}'$) are pure spinors just as ($\la$, $\bar{\la}$) respectively. Since the operator diverges no more than is taken care of by $[\mathrm{d}f][\mathrm{d}\bar{f}]$, it would not contribute any negative factors of $\xi$ or $\sigma$ to an integrand. \cite{Berkovits&Nekrasov}

For such a construction to be possible, suitable ``translation''
operators $W_\alpha$
and $\bar{W}_\alpha$ must be available, and the process must be
possible to describe through the introduction of a new regulator. To
begin with the operators, the $W$'s represent gauge invariant versions
of the $\omega$'s. If the $\omega$'s were to be used instead of the
$W$'s, eq. \eqref{eq.GuessOp} would incorporate $\la'=\la+\varepsilon
f$, which is not necessarily a pure spinor. For that to be true, a
certain gauge must be chosen, with specific constraints on
combinations of for example $\omega$ and $\bar{f}$, but the equations
need to be independent of the choice of gauge. That is what calls for
the construction of  derivatives that are obtained using
projectors on the cotangent
space of pure spinors in $D=11$:
\begin{equation}\label{eq.projector}
W_\alpha= {P^\beta}_\alpha \omega_\beta: \qquad P=\mathbb{I}-P_\perp
\end{equation}
The reason for this expression for the projection ${P^\alpha}_\beta$
is that it acts as the  identity on tangent vectors $d\la^\beta$ which
fulfil $(\la\g^ad\la)=0$. Essentially, the pure spinor constraint is
enforced by the removal of the parts which would not automatically
obey it, by the presence of $P_\perp$. The construction of $W$ is
far more complicated in $D=11$ than in $D=10$.
It is presented in detail in appendix \ref{app.Pro}:
\begin{align}
P^\alpha{}_\beta&=\delta^\alpha{}_\beta
-\frac{1}{4\alpha}(\g_a \fb)^\alpha(\la\g^a)_\beta
-\frac{1}{2\alpha\beta}(\g_a \fb)^\alpha(\la\g^{ai}\la)(\fb\g_{bi}\fb)(\la\g^b)_\beta+\nonumber\\
&\quad
+\frac{3}{4\alpha\beta}(\g_{[ij}\fb)^\alpha(\fb\g_{kl]}\fb)(\la\g^{ij}\la)(\la\g^{kl})_\beta
\end{align}
Here, $\alpha=(\la \fb)$ and
$\beta=(\la\g_{ab}\la)(\fb\g^{ab}\fb)$.
Moreover, with a short notation according to
$P_\perp^\alpha{}_\beta=(\g_a\fb)^\alpha
R^a{}_b(\la,\fb)(\la\g^b)_\beta$, the above gives at hand a
``translated'' pure spinor:
\begin{equation}
\la'=e^{(\e W)}\cdot\la^\alpha=\la^\alpha+\e^\alpha-\frac12(\g_a\fb)^\alpha R^a{}_b(\la+\e,\fb)
((\la+\e)\g^b(\la+\e))
\end{equation}

Analogously, the form for $\bar{W}$ is found by a change of all barred
and unbarred variables:  $\la \leftrightarrow \lb$ and correspondingly
for all other indices which exist in both versions
\cite{Berkovits&Nekrasov}. What is important for the continuation of
the description of the regularisation procedure is that the
correct, gauge invariant derivatives can be constructed.

In order to describe the process through the introduction of a new regulator, a full set of variables must be introduced. Not only the pure spinors are needed, but also constant fermions $g_\alpha$ and $\bar{g}_\alpha$ with constraints: \cite{Berkovits&Nekrasov}
\begin{equation}
g\g^af=\bar{g}\g^a\bar{f}=0\li, \quad [Q,f]=\{Q,\bar{g}\}=0\li, \quad [Q,\bar{f}]=\bar{g}\li, \quad \{Q,g\}=f
\end{equation}
This can be compared to the introduction of the non-minimal variables,
though the unbarred variables imitate ($\la$, $\theta$)
instead. With\footnote{Here $\bar{\omega}\rightarrow s$ denotes that
  every $\bar{\omega}$ in the expression is replaced by an $s$.}
$V=\bar{W}(\bar{\omega}\rightarrow s)$ we then have that
$\{Q,gW+\bar{f}V\}$ contains a term $fW+\bar{f}\bar{W}$, as well as
that $\{Q,\bar{f}g\}$ contains a term $f\bar{f}+g\bar{g}$, which means
that it is possible to do a construction as in
eq. \eqref{eq.GuessOp}. In fact, the correct regularisation for
an operator (acting on fields etc.) is: \cite{Berkovits&Nekrasov}
\begin{align}\label{eq.regGen2}
O(z)\rightarrow\int[\mathrm{d}f][\mathrm{d}\bar{f}][\mathrm{d}g]
[\mathrm{d}\bar{g}]e^{-\{Q,\bar{f}g\}}
e^{i\varepsilon\{Q,gW+\bar{f}\bar{V}\}}O(z)e^{-i\varepsilon\{Q,gW+\bar{f}\bar{V}\}}
\end{align}
This takes care of the divergences of the operator in question, if the
singularity is not too bad (see below). It also performs another regularisation,  necessary due to the
introduction of the integration over the new variables: the first type
of regularisation discussed above, for the set of variables
represented by ($f$, $\bar{f}$, $g$, $\bar{g}$).

In total, we have a procedure that provides a smoothing of a singular
operator. Performed once, it takes care of singularities within the
limits of eq. \eqref{eq.reqSG}. If that is not sufficient, the
procedure can be performed over and over and again. Each time, new
variables are introduced and the smoothing of the operator or
integrand corresponds to a smoothing of the maximally allowed
singularity in  eq. \eqref{eq.reqSG}. Eventually, this gives at hand a
regular expression. \cite{Berkovits&Nekrasov}

The number of times this procedure must be performed is roughly
determined by how high a negative power of $\eta$ there is present in
the expression in question, as it is the only scalar that shows up
with negative powers in the operators. As such, the power of $\sigma$
represents the worst singularity, with ``one smoothing procedure''
taking care of no more than $\eta^{-(11/2)}$. A general
regularisation is therefore typically very intricate. The
geometrical interpretation though, can be compared to a ``heat kernel'' regularisation
\cite{Berkovits&Nekrasov}. It can most certainly be replaced by
  a more geometric construction explicitly involving smoothing by
the Laplacian on
  the pure spinor space. This would involve further geometric
  investigations along the lines of ref. \cite{CederwallGeometry}, but it
  is not clear that it would provide any calculational advantage,
  especially since it would not respect the splitting into separate
  holomorphic and antiholomorphic factors of the present approach.

\subsubsection{The interpretation of regularised amplitudes}\label{subsubsec.intGenRegAmp}
The above establishes that the proposed regularisations
work. But in what way do they show up in a typical integrand,
originating in e.g. an amplitude diagram?

Worth to note about regularisations in general is that they can
be performed in random order and at any time up until the properties
which they alter are being used. For the general regularisations
described above, this is not much of a restriction, since they do not
meddle with an expression very much, in contrast to the loop
regularisation which will be introduced below. The regularisation in
eq. \eqref{eq.Genreg1} contains no derivatives, and therefore
changes none of the variables that were present
before the introduction of the regulator, while the
variables the regularisation method described in
eq. \eqref{eq.regGen2} acts on are
replaced by variables of very similar (indeed, partly identical)
properties. Furthermore, the introduced variables exist in a very
limited region (one operator) and are integrated out, properties which
combine in such a way as to restrict the impact of the alteration to
one thing only: the regularisation of the expression in
question.

In specific, this means that the general regularisations can be
inserted at any time up until the integration over the superspace
variables, yielding the same result. As such, the
regularisations due to the pure spinors can be kept implicit in
the examinations of the amplitude diagrams, and as a rule they are
never mentioned.

However, not all regulators are as well behaved as the above, which
will be noted in the next subsection. The regularisation which
is necessary to introduce due to the presence of loops acts on the
variables in a slightly different way, exchanging one for another. In
contrast to the regularisations above, that means that the
regularisations must be performed at a certain time --- before
properties of different variables are being used.

\subsection{Loop regularisation}\label{subsec.loopReg}
At the introduction of loops, in contrast to
a description with tree
diagrams only, a problem with the phase space field theory
construction typically occurs: it lacks the description of the freedom
of momenta etc. that is introduced by the formation of a loop. This
occurs when the propagator is too local with respect to one of its
bosonic variables, e.g. a function $\delta(x-y)$ in space. At the
integration over the vertices attached to a loop, compare figure
\ref{simpAmplDiags}, the last integration then must be performed over
$\delta(0)$, which diverges. This last expression, present when one
integration remains, is what is interesting in the examination of
amplitude diagrams, and something divergent does not make any sense.

Ordinarily, this problem does not show up due to the presence of a
kinetic term in the theory. As such the propagator in the momentum space is expressible in terms of an
exponential of the Laplacian, instead of a factor 1, and the phase space propagator is not too local, but a Gaussian curve:
\begin{equation}\label{eq.LaGa}
\frac1{p^2}\sim\int_0^{\infty} e^{-ap^2}\rd a \quad\leftrightarrow \quad e^{-a(x-y)^2}, \quad a>0
\end{equation}

The above is also the case for the variable $x$ in maximal supergravity described with the pure spinor formalism, but only
because of the gauge fixing, which gives at hand the relevant
dynamics ($b/p^2$). The theory contains two more bosonic variables $(\la,\lb)$
without corresponding kinetic terms. Moreover, the propagator is too
local with respect to them, represented by $\delta$-functions, and
needs to be changed into a less local one in order for the theory to
make sense. One possible way to do that is to deform the theory by
introducing suitable kinetic terms, giving at hand properties for the
propagator as in eq. \eqref{eq.LaGa} with respect to the pure spinors
as well as $x$. However, such a deformation is not desirable, as the
theory would be deformed as well, and there would be no way of
connecting results provided by it to results in maximal supergravity.

Instead, the key feature to be recognised is that the pure
spinor formalism gives at hand a possibility to \emph{imitate} the
construction in eq. \eqref{eq.LaGa}, through the use of
BRST-equivalent terms. Either a suitable regulator is added, a process
which bears a striking similarity to the expression on the left hand
side in eq. \eqref{eq.LaGa}, or else the $b$ operator might be changed
into something less local, yet still with the property
$\{Q,b\}=\partial^2$. Such a conversion, in either direction, can of
course be done for any number of propagators in a loop diagram. In
essence, it means that it is possible to exchange the too local
propagator in phase space for a less local one. The result is a
remaining integration over a function which is finite, and the problem
is solved.

A possible observation at this point is that the general
regularisation described in eq. \eqref{eq.regGen2} performed on
a $b$ operator might do for the description of a not too local
propagator. However, that regularisation is quite intricate and
it is far from obvious how to extract any loop properties with such a
construction. At the end of the day, another solution is sought for,
with which properties of amplitude diagrams might be observed before
the final regularisation in eq. \eqref{eq.regGen2}.

A $Q$-invariant alteration of the $b$ operator into something less
local, still obeying $\{Q,b\}=\partial^2$, might be a convenient way
of finding a solution. The construction of eq. \eqref{eq.regGen2} with $O\rightarrow b$
  seems to provide such a deformation, but is extremely unpractical to
use in loop calculations. We have put some effort into trying to find
alternative explicit expressions for $b$, not involving extra
variables, but have not succeeded. Although we still do not want to exclude the
possibility of finding a more efficient, non-local but regular, propagator; given the present status of the
propagator, what remains is the introduction of one
more regulator.

\subsubsection{Loop regularisation in superspace}
In superspace, the only apparent way of regularising a loop,
while retaining any possibility of extracting results, is to
recognise the freedoms of momenta etc. in the loop, instead of
using the last $\delta$-part of the propagator, in each loop, which
would diverge. In figure \ref{vertices} the latter corresponds to
gluing two connected propagators together, so that there only is one
$\delta$-function which by the fusion turns into ``$\delta(0)$''. But
instead of using this last $\delta$-function, which is superfluous and
does not capture the loop properties, all the momenta in the loop (of
number $I$) are changed so that the freedoms of the loop momenta are
described: $\partial \rightarrow
\partial+\partial^I$ for the terms in the loop in consideration, and
the corresponding for $D^I$, $N^I$, $N_{ab}^I$, $\bar{N}^I$,
$\bar{N}_{ab}^I$, $S^I$ and $S_{ab}^I$. Each of these ``coordinates'' of
the cotangent space represents a new variable, ranging in value within
its domain, so it is also necessary to introduce the corresponding
integration measures, which implies that the process corresponds to
the following: \cite{JB,Berkovits&Nekrasov}
\begin{equation}
\int f(b,p)\delta(0)[\mathrm{d}Z]\rightarrow \int f(b_{new},p+p^I)[\mathrm{d}Z]\mathrm{d}^Dp^I[\mathrm{d}D^I][\mathrm{d}N^I][\mathrm{d}\bar{N}^I][\mathrm{d}S^I]
\end{equation}
Here, $b_{new}$ is such that each $b$ depends on the momenta and the
loop momenta of the loop that the propagator is part of\footnote{All
  loops need to be regularised, so when a propagator is shared
  between different loops, the $b$-ghost depends on the original
  momenta and the momenta of the loops: $\partial \rightarrow \partial
  +\sum_I \partial^I$.} \cite{JB}. The last three integration measures
are schematically given as\footnote{$N$ here temporarily
denotes both $N$ and $N_{ab}$, and
  the initial integration is with respect to $\omega$, which however
  must show up in the gauge invariant quantities described by the
  $N$'s. The corresponding is true for $\bar{N}$ and $S$.}:
\begin{align}
[\mathrm{d}N^I]&\sim\la^{-16}\mathrm{d}^{23}N^I\li,\quad
[\mathrm{d}\bar{N}^I]\sim\bar{\la}^{-16}\mathrm{d}^{23}\bar{N}^I
\li,\quad
[\mathrm{d}S^I]\sim\bar{\la}^{16}\mathrm{d}^{23}S^I\label{eq.dNdS}
\end{align}
The powers of $\la$ and $\lb$, which are only given schematically,
follow from the simple observation that while
$[\mathrm{d}\la]\sim\la^{-7}\mathrm{d}^{23}\la$, the inverse metric governs the
integration over cotangent space, which then becomes
$[\mathrm{d}N]\sim\la^7\mathrm{d}^{23}\omega\sim\la^{-16}\mathrm{d}^{23}N$.

However, the amplitudes are independent of $S^I$, $S_{ab}^I$,
$\bar{N}^I$ and $\bar{N}_{ab}^I$, the forms of which are described in
eq. \eqref{eq.gaugeN}, as these quantities do not show up in any
expression. They certainly are not present in any term which is not a
regulator, nor in the regularisation presented in
eq. \eqref{eq.regGen2} due to the form of it. Moreover, the absence of
the two last quantities gives at hand that the amplitudes cannot
depend on $N_{ab}^I$ or $N^I$ either \cite{JB}, so a regulator is
needed in order for the integration to make sense, $e^{\{Q,\chi\}}$
\cite{JB,Berkovits&Nekrasov}:
\begin{align}
\{Q,\chi\}&=[\chi=-k(NS+N_{ab}S^{ab}),\quad k>0]=\nonumber\\&=k\Big((\la D)S+(\la\sigma_{ab} D) S^{ab}-N\bar{N}-N_{ab}\bar{N}^{ab}\Big)
\end{align}
Note that the components of $D^I$ in each $D$ etc. here have been kept implicit, as is most convenient. As long as the free loop momenta have not been integrated out, they are present in every momentum term.

The expression above brings us back to the interpretation of the
regularisation as originating in the introduction of a
regulator. It has certain properties which will be discussed in more
detail below, such that an $r$ can be turned into ``$\la D\bar{\la}$''
for example \cite{JB}, but for now, we continue the examination of the
impact of the regulator on the newly introduced integrations. As it provides the only terms that the integrations in
eq. (\ref{eq.dNdS}) concern, those integrations (with the
regulator and all other terms implicit) give at hand \cite{JB}:
\begin{equation}\label{eq.regpart1}
\int[\mathrm{d}N^I][\mathrm{d}\bar{N}^I][\mathrm{d}S^I]
\sim\frac{1}{\la^{16}\lb^{16}}\int[\mathrm{d}S^I]
\sim \la^7 (D+\sum_I D^I)^{23}
\end{equation}
From this we can see, for example, that in order for the free,
fermionic momenta of $S$ to be integrated out, the terms need to be
taken from the regulator, which brings down an extra term ``$\la D$''
for each $S$ that needs to be integrated out. The end result is
furthermore quite convenient as a factor $\la^7$ has been provided for
the volume form $[\mathrm{d}\la]\sim\la^{-7}\mathrm{d}^{23}\la$ in
$[\mathrm{d}Z]$. However, as there are 32 spinors $D^I$ for each loop,
the result for the entirety of the newly introduced integrations is:
\begin{equation}\label{eq.FInteg}
[\mathrm{d}N^I][\mathrm{d}\bar{N}^I][\mathrm{d}S^I][\mathrm{d}D^I]\mathrm{d}^Dp^I\stackrel{\eqref{eq.regpart1}}{\sim} \la^7 \mathrm{d}^{9}D^I\li\mathrm{d}^Dp^I
\end{equation}

The effective integration thus requires 9 $D^I$'s per loop to be
extracted from the amplitude diagrams in order for a non-zero
result. These must be provided either through the $D^I$'s already
present in the propagators and vertices, or through the use of the
regulator.

The last part of eq. \eqref{eq.FInteg} is the integration over the
momenta $p^I$ present in the loops, which is done up to some momentum
cut-off limit, $\La$. 
With respect to the finiteness of the
theory, these last integrations need further examination. Part of
this, more specifically the UV behaviour of the amplitude diagrams,
will be discussed in the next section.

All in all, the introduction of the momenta freedoms and their
integration measures changes the integration over each loop so that a
finite result can be obtained. The procedure corresponds to an
integration over the cotangent space to a point in superspace,
described with the real metric, which is Calabi--Yau.

\subsubsection{The matter of loop regularisation}\label{subsub.matterLoop}
As noted for the general regularisations, regulators can be introduced
randomly up until their properties, or the properties they affect, are
used. The latter is what makes the loop regularisation a bit trickier
than the general regularisations.

The apparent condition brought on by the regularisation is that 9 $D^I$'s per loop need to be extracted from the integrand. This in itself is rather harmless, though crucial for the amplitude properties. The complication is brought on by that $r$ can be converted into $D$:
\begin{equation}\label{eq.r2D}
r_\alpha \rightarrow
\left\{\begin{array}{l}
(\la D)\lb_\alpha\\
(\la \g_{ab}D)(\lb\g^{ab})_\alpha
\end{array}\right.
\end{equation}
The above calls for two different observations. Firstly, the first
transformation cannot be applied to any $r$ which is a part of a
structure as the one described in eq. \eqref{eq.L}. Due to symmetry
properties, the result would be zero. That is why the first
transformation only can be applied to $r$'s part of the operator $T$
in the 4-point vertex, and only after the second transformation has
been used at least once, at that.

Secondly, the above clearly changes the algebraic properties of the
indices attached to the $r$'s which are to be transformed. Unless the
indices, naively described as $r_\alpha r_{\beta}$ ``before''
regularisation, are part of structures that ensure them to be
antisymmetrised regardless of the presence of the $r$'s, such as
individual operators, they are not necessarily restricted to be
antisymmetrised. Then, one has to take the regularisation into
consideration before using the properties of the non-minimal variable
$r$. In short, the constituent part need to be regularised before they
are put together.

The simplest solution to the above might have been not to use the
property of the regulator which changes variables according to
eq. \eqref{eq.r2D}. Indeed, where that is possible, that is what we
will do. It is, as described above, quite all right to choose to
examine any term in the expansion of a regulator (as long as the
constant one is kept, of course); the answer will be correct, provided
that it is not nonsense like $0\times\infty$. However, without the use
of the regulator, all terms with a factor $r^{x}$ with $x>23$ would
incorrectly give at hand a zero contribution, due to the fact that one
$\lb$ invariably is part of the integrand. The condition $(\lb\g^a
r)=0$ then reduces the degrees of freedom of the fermionic spinor $r$
to 23. But the fact that the $r$ cannot show up in any higher powers
than 23 does not prevent the presence of ``converted'' $r$'s. So all
terms with $r$ to a higher power than 23 need to be brought into a
form with $r^{23}$ in order to be interpreted correctly, a procedure
which is carried through with the help of the loop regulator.

To be precise, one may note that $r$ and $D$ in a certain manner can
be treated on an equivalent basis. Not only due to the transformation
property of eq. \eqref{eq.r2D}, but due to the fact that the presence
of an $r$ in an operator only comes at the expense of a $D$, compared
to the terms with lower powers of $r$, compare
e.g. eq. \eqref{eq.b}. At the examination of whether or not an
expression needs to be regularised with respect to $r$, one
might therefore consider only the terms solely consisting of $r$'s,
without the presence of any $D$, and note that a regularisation with
respect to $r$ only gives a non-zero result if the power of $r$
present exceeds or equals $23+9L$. This allows for the $r^{23}$
remaining after the conversion and the $(D^I)^9$ required for each
loop integration in order for a non-zero result, and provides a most
convenient way of determining when there might be too many $r$'s
present in an expression.

Moreover, note that the $r$'s remaining after the loop
regularisation are distributed across the former positions of
$r$. Where they sit does not alter any properties of the
expression. This is on equal footing with the fact that if part of a
loop structure, which on its own does not need to be regularised,
displays certain properties when it is not regularised, those
same properties will be present when it is regularised as a part
of the entire loop structure. We have BRST invariance, so the
properties obtained are the correct ones. The converted $r$'s need
only be taken into account when properties dependent on structures
with too many $r$'s present are being examined.

\subsection{Amplitude characteristics}
The loop regularisation and the formation of loops are the key
features restricting the constituent parts of the loops and in what
ways they can be put together into amplitude diagrams with tree
parts. The first condition, 9 $D$'s to be extracted from each loop, is
fairly straightforward. The conversion of $r$'s into $D$'s and the
fact that loop operators are allowed to act in all directions on the
other hand unsurprisingly gives at hand that even the description of a
diagram with only a few loops is quite intricate. The most apparent
properties will be discussed below though.

To begin with, we may take note of the different parts of an amplitude
diagram. There are two different kinds: planar and non-planar
diagrams, the former which can be drawn on a plane without ever
crossing a line. That is a property which refers to the composition of
the loop structure. In the general scheme of things though, tree parts
are connected to a loop structure via a number of so called outer
legs, as are the external legs that do not connect to tree
parts. Moreover, although some diagrams, tree diagrams, do not contain
a loop structure, those will not be treated below, as no surprising
restriction on the properties thereof occurs.

What we will begin by discussing is the loop structure itself,
starting out with the most simple case, the one-loop amplitude. An
examination of it shows why bubbles and triangles are absent in
amplitude diagrams. This naturally carries on towards a discussion on
the splitting of momenta between different loops, particularly with
respect to the presence of outer legs, and to some degree the
conditions on non-planar diagrams also. It is, however, difficult to
discern properties of structures which require regularisation with
respect to $r$. The question of when such regularisations are
required is addressed in the next section, on the UV characteristics
of the amplitude diagrams.

Subsequently, we discuss the picture of the loop structures which has
emerged, and finally, the general conditions on the amplitudes are
taken into account, with respect to the ``outer components'' that are
not part of the loop structure. In total, this gives at hand a fairly
detailed, general description of the amplitude diagrams.

\subsubsection{No bubbles or triangles}\label{noBubbles}
A known feature of amplitude diagrams in maximal supergravites is that they cannot contain any bubbles or triangles
(loops with less than four vertices attached) \cite{JB,J1,J2,J3}. In
the formalism here presented, this is a consequence of the fact
that each loop
for a non-zero result needs to contribute 9 $D$'s to the integration
over the loop variables. As is shown below, this cannot be provided by
a bubble or a triangle, no matter what their outer legs are connected
to, so no bubbles or triangles can exist as parts
  of a loop structure.

The reasoning goes as follows. To begin with we address the question
of whether or not loop regularisation with respect to $r$ will be
necessary, compare subsection \ref{subsub.matterLoop}, which it turns
out not to be. Each set of vertex and propagator operators\footnote{A
  set of vertex and propagator operators here denotes the presence of
  one vertex and one propagator, i.e., two $R^a$'s and one $b$ in the
  case of a 3-point vertex, and the additional $T$ in the case of a
  4-point vertex. A bubble consists of no more than two such sets and
  a triangle of three, although the 4-point vertex only can show up
  once.} contains at most $r^7$, or $r^{10}$ (for the 4-point vertex),
yielding that the highest possible power of $r$ present in a one-loop
amplitude with less than 4 vertices is $r^{24}$. Although this power
is greater than $23$, that term is accompanied by no $D$ and cannot
give at hand $r^{23}D^9$ from regularisation, and as such
vanishes. Effectively, a bubble or a triangle contains no terms with
$r^x$, $x>23$ on its own. Thus, as $e^{\{Q,\chi\}}$ is $Q$-exact, the
exponential term does not determine whether or not a bubble or a
triangle exists, and it may be treated as a factor 1. As such, $r$'s
cannot be converted into $D$'s, and a propagator or a vertex can at
most contribute with two $D$'s to the loop integration.

This would be enough to exclude bubbles, but not triangles. However,
each unregularised vertex cannot contribute with more than one $D$ to
one and the same loop integration, since the antisymmetrisation of the
indices in $R^a$ and $R^b$ ensures that the indices of their two $D$'s
are symmetrised, whereas they need to be the opposite of
that. Furthermore, the part of the propagator which might contribute
with two $D$'s looks like $f(\la,\bar{\la},r)^{ijk} D\g_{ijk} D$,
compare eq. \eqref{eq.b}. This quantity is fermionic, which gives at
hand that in order for two propagators to both contribute with two
$D$'s to the loop integration, those two parts must be be
antisymmetrised. At the same time, the four $D$'s must be
antisymmetrised, as the 9 $D$'s are. In order for the latter condition
to be true, the two factors of $D\g_{ijk} D$ need to be symmetrised,
in which case the former condition is such that the two factors of
$f(\la,\bar{\la},r)^{ijk}$ need to be antisymmetrised. This is an
apparent contradiction as the sets $ijk$ and $i'j'k'$ cannot be both
symmetrised and antisymmetrised at the same time, with the consequence
that two propagators cannot both contribute with two $D$'s to one and
the same loop integration.

In total, only one propagator can contribute with $D^2$ to one and the
same loop integration, the rest of the propagators as well as the
vertices cannot contribute with more than one $D$ each. This renders
bubbles non-existent, since they at most can contribute $D^5$ to the
loop integration. The same goes for the triangle, which at most can
contribute $D^7$ to the loop integration. For 9 $D$'s, at least four
legs need to be attached to the loop.

Any one-loop amplitude thus must have at least four external
legs. Moreover, with the analogy observed in subsection
\ref{subsub.matterLoop}, every loop in an amplitude diagram must have
at least four vertices, so neither bubbles nor triangles exist in maximal supergravity.

\subsubsection{The splitting of momenta between loops}
It is important to note, that the absence of bubbles and triangles not always is enough to render an amplitude diagram non-zero, with respect
to the 9 $D$'s that need to be claimed for each loop. For example, if
a propagator is shared between several loops (possibly more than two
if the diagram is non-planar), its (absolute) maximum contribution of
3 $D$'s to the surrounding loops might fall short of what is
needed. In such a case, there would be a requirement for one or
several of the outer legs to be attached to some of the loops in
question in order for the diagram to be non-zero. As each outer leg
adds a propagator to the loop to which it is attached, the $D$'s that
propagator could provide to the loop integration would ease the
requirement on the contribution from the first propagator, otherwise
shared between too many loops.

This brings us to a more detailed examination of how the momenta in a
loop can take values in the degrees of freedom introduced by the
loops, as well as the original momenta, which is another feature of
the loop regularisation. This is represented by
e.g. $\partial\rightarrow \partial+\sum_I\partial^I$ with $I$ summed
over the different loops the original momenta constitute a part of. To
simply assume that these types of momenta end up distributed randomly,
even taking into account that 9 $D^I$'s must be removed for each loop
integration, is to pursue matters a bit too far though. Some terms
vanish.

Consider, for example, the case of a one-loop amplitude with $N$ outer
legs, or any part of a loop which can be constructed from one
propagator to which $N-1$ outer vertices are attached. It has $N$
propagators that all, naively, have momenta:
$\partial+\partial^I$. They are all a part of the same loop, and no
other, so even the possible term $\sum_J \partial^J$, for the general
case, can just as well can be regarded as a $\partial^I$. However, a
$b$ operator placed between two certain vertices, called number 1 and
2, can act in any direction since it is a part of a loop. The way it
acts on or is acted on by the amplitude diagram and its loop
regularisation is distributed across the entire structure. In fact,
the description is equivalent to one where that $b$ operator is not
placed between vertex number 1 and 2, but instead placed on any of the
other propagators, right next to the $b$ operator which was there in
the initial description. This is perfectly fine\footnote{Note that
  this is not equal to having two $b$ operators on the same propagator
  in the picture where the loop momenta are absent. Such a term would
  yield $\{b,b\}=0$. However, this is not true when the loop momenta
  are present, and thus $b$ operators can be brought onto the same
  propagator through partial integration, with a non-zero result.}. Moreover, the two
terms contain at the most $r^6$, so the ensuing structure can be
analysed  before any regularisation with respect to $r$. For example,
some combinations of those two $b$ operators depend entirely on either
the momenta $\partial$, or the loop momenta $\partial^I$, in which
case we have a factor $\{b,b\}=0$ so that the term vanishes.

Let us be explicit. If an operator $D\g^{abc}D$ were to be part of
\emph{one} loop regularisation, it would effectively contain
$D\g^{abc}D+2D^I\g^{abc}D+D^I\g^{abc}D^I$. This happens to the $b$
operators as well, and because of the property $\{b,b\}=0$, two
propagators which are part of the same loop structure cannot be
dependent entirely on the same set of momenta $\partial^I$, with $I$
taking the value of a loop, or not. Most propagators are split
between the momenta of different loops, as the term $D^I\g^{abc}D$ is.

In specific, in the setting above, only one of the $b$ operators can
depend entirely on the loop momenta $\partial^I$. The rest at the very
least contains one $D$ each, a property which will come in handy in
the next section, on the UV behaviour of amplitude diagrams. In
relation to the previous subsection, this is also the fundamental
property that explains why no more than one of the propagators in a
bubble or triangle can contribute with two $D$'s to the loop
integration.

Moreover, the examination can be drawn a bit further. We note that the
number of propagators that can be compared in this way, at the same
time, without any transformation of $r$, is at most 7. Out of these,
one may depend on $\partial^I$, another on $\partial$, but the rest
must be split between the two sets of momenta, and the entire
expression for $\{b(\partial+\partial^I),b(\partial+\partial^I)\}$ can
be investigated. It is non-zero, but only for terms that in
combination have $r^3$ as a part of them. In specific, the remaining
components originate in $b_1$ and $b_2$ and consists of terms with
$D^3$ split between the momenta. When $r$ is not regularised, the
propagators form no terms with $\partial^2$.

The vertex operators $R^a$ can be considered in a similar way. As they
are bosonic and act around a loop back to its initial position,
without vanishing along the way, the part which acts on components of
the loop may be considered to give a zero contribution. The $R^a$ contributes its
components to the loop regularisation, either through $D^I$ or
$\partial^I$. The latter cannot form $(\partial^I)^2$ with another
$\partial^I$ coming from vertex operators or propagators without any
$r$ transformed, though. Effectively, this brings with it that no
$(\partial^I)^2$ can be formed, in total, and one $R^a$ for each
vertex must contribute $D^I$ to the loop integration, for diagrams
that need not be regularised with respect to $r$, which will be
discussed further in section \ref{sec.UV}. For a summary of what all
this means with respect to actual loop components, we now turn to the
most simple example possible.

\subsubsection*{Illustrative example: the 4-point one-loop amplitude}
The simplest case there is for a loop diagram is the one-loop
amplitude which consists of four propagators and four 3-point
vertices. In order to illustrate the properties presented above, we
will have a closer look at the non-zero components of it.

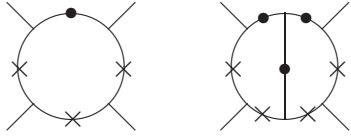
\begin{figure}
\begin{picture}(250,48)(-110,0)
\put(24,24){\circle{38}}
\put(0,0){\line(1,1){10}}
\put(38,38){\line(1,1){10}}
\put(0,48){\line(1,-1){10}}
\put(38,10){\line(1,-1){10}}
\put(24,44){\circle*{4}}
\put(20,1){$\times$}\put(0,20){$\times$}\put(39,20){$\times$}
\put(104,24){\circle{38}}
\put(80,0){\line(1,1){10}}
\put(118,38){\line(1,1){10}}
\put(80,48){\line(1,-1){10}}
\put(118,10){\line(1,-1){10}}
\put(104,4){\line(0,1){40}}
\put(96,42){\circle*{4}}\put(112,42){\circle*{4}} \put(104,23){\circle*{4}}
\put(80,20){$\times$}\put(119,20){$\times$}\put(91,3){$\times$}\put(108,3){$\times$}
\end{picture}\caption{The distribution of terms on the one-loop
  diagram, as well as a possible distribution for the two-loop
  diagram. The filled dots represent $b_1$ terms with $D^2$ and the
  $\times$'s terms of $b$ that only contribute with one $D$ to the
  loop integration(s), just as each external vertex does. The internal
  vertices each contribute one $D$ to both loops.\label{fig.loopsYM}}
\end{figure}

It contains at most a power of $r^{4\times7}=r^{28}$ where no $D$ is
present and the presence of a $D$ comes at the expense of one $r$,
yielding that all terms with $r^x$, $x>23$ vanish. Therefore, just
as for bubbles and triangles, the conversion between $r$ and $D$ need
not be considered and the maximal contribution of $D^I$ to the loop
regularisation is 9, the minimum required. This happens when one
propagator contributes with $(D^I)^2$ through the presence of a term
$b_1$, compare subsection \ref{sub.Op}, and the rest of the vertices and propagators contribute with one $D^I$
each through the presence of four $R_1$ and three $b_0+b_1+b_2$, the
latter which need to be split between momenta (apart from $b_2$) so
that they contain one $D^I$ each. Moreover, due to the lack of the
$r^0$ term in the antisymmetrisation of the split $b$ operators, as
mentioned right before this example, not two of the propagators can be
represented by $b_0$. Compare figure \ref{fig.loopsYM}.
\\
\\
To conclude, it is appropriate to comment on the change of this once
several loops are taken into account, as also is illustrated for a
simple example in figure \ref{fig.loopsYM}. It is not possible to
shift a $b$ operator onto a propagator with different loop variables
from the ones in the operator itself, but some relations can be
deduced even so. Within each loop, the rules above apply. For example,
for the two-loop diagram in figure \ref{fig.loopsYM}, which does not
need to be regularised with respect to $r$, at most three propagators
can contribute $D^2$ to the loop integrations, one for each loop and
one split between the loops, placed on the propagator they share. The
corresponding is true for the vertex operators as well.

Furthermore, the property that no term with $(\partial^I)^2$ can be
formed applies for all loop structures that have not been regularised
with respect to $r$, even the ones with multiple loops. This however
comes very close to having less to do with
general properties than the minimal
structure. For example, at least one $D$ is required from each operator
(acting into the loop) for the amplitude to be non-zero. This will in
part be discussed a bit in the next subsection, and furthermore in the
next section, on the UV properties of the amplitude diagrams, where we
examine what is required of a loop structure in order for it to need
to be regularised with respect to $r$.

Generally, it is difficult to discern properties of multi-loop
amplitudes where $r$ is regularised. The descriptions simply allow for
a lot of freedom as to what the variables are and how they may act,
which is in need of further examination.

\subsubsection{The structure of loop diagrams}\label{subsub.loopstr}
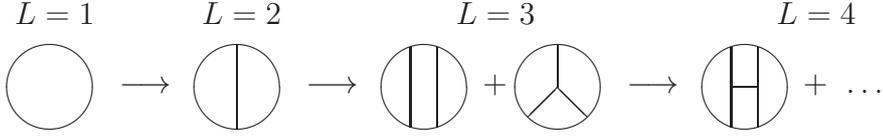
\begin{figure}
\begin{picture}(380,50)(45,6)
\put(70,24){\circle{30}}\put(57,48){$L=1$}
\put(96,22){$\longrightarrow$}
\put(140,24){\circle{30}}\put(140,8){\line(0,1){32}}\put(127,48){$L=2$}
\put(166,22){$\longrightarrow$}
\put(210,24){\circle{30}}\put(205,9){\line(0,1){30}}\put(215,9){\line(0,1){30}}
\put(232,22){$+$}\put(222,48){$L=3$}
\put(260,24){\circle{30}}\put(260,24){\line(0,1){16}}\put(260,24){\line(1,-1){11}}\put(260,24){\line(-1,-1){11}}
\put(286,22){$\longrightarrow$}
\put(330,24){\circle{30}}\put(325,9){\line(0,1){30}}\put(335,9){\line(0,1){30}}\put(325,24){\line(1,0){10}}
\put(352,22){$+$}\put(368,22){$\ldots$}\put(342,48){$L=4$}
\end{picture}\caption{Illustration of how the number of propagators in
  a loop structure without external legs goes as $3L-3$ for $L\geq2$
  when the 4-point vertex is absent, while the number of vertices goes
  as $2L-2$. It also shows that the minimal amount of outer legs
  necessary in order to avoid bubbles and triangles for all diagrams
  at a certain loop order is $4$ for $L\leq2$, $3$ for $L=3$ and no
  more than $2$ for $L\geq4$, if there is no 4-point vertex
  present. Otherwise, the minimal $j$ need to be increased by one
  outer leg, and $j=2$ is not reached until
  $L=5$.\label{fig.NumLoops}}
\end{figure}

We now turn to a general inventory of the components of a loop
structure with $L$ loops, which e.g. will be of use in the next
section when the question of when it is necessary to do a
regularisation with respect to $r$ is addressed. Temporarily
disregarding the one-loop, we may observe that loop structures built
from only 3-point vertices and without outer legs connected to them
contain $3(L-1)$ propagators and $2(L-1)$ vertices, compare figure
\ref{fig.NumLoops}. To go from a structure with $L$ loops to a
structure with one loop more, a propagator must be added to the
diagram, and its two ends connected to already existing propagators,
that each become divided in two, yielding a total of three propagators
and two vertices added to the diagram with $L$ loops. The total number
of propagators in a $L$-loop structure thus is $3(L-1)+j$, with $j$
the number of outer legs. This is at least four for $L\leq2$, in order
to avoid bubbles and triangles. For $L=3$ at least 3 outer legs are
necessary in order to get one non-zero amplitude, and for $L\geq4$
there are amplitudes that are non-zero for the minimal
requirement\footnote{The reason for this will be explained in the next
  subsection.} of $j=2$. The total number of vertices is $2(L-1)+j$.

The above is, of course, true only in the absence of the 4-point
vertex, which however can show up only once in an amplitude, as
described in subsection \ref{subsub.vert}. When this special case
occurs, the 4-point vertex can be regarded as having taken the place
of a 4-point tree part as the one shown in figure
\ref{simpAmplDiags}. By this analogy, two 3-point vertices and one
propagator are turned into the 4-point vertex, giving at hand the
presence of one 4-point vertex, $3(L-1)+j-1$ propagators and
$2(L-2)+j$ 3-point vertices.

\subsubsection{The importance of components outside a loop structure}
A loop may appear alone, only accompanied by external fields, or exist
as a part of a loop structure, i.e., several loops that are connected
by propagators. Such a structure possibly has tree diagrams extending
from it, resulting in external fields. No matter what the appearance,
at least 4 external fields are necessary for an amplitude diagram to
exist \cite{JB}, and at least two legs need to be connected to the
loop structure, as mentioned above.

The latter occurs since vacuum amplitudes (no outer legs connected to
the loop structure) or amplitudes with no more than one external leg
cannot occur \cite{JB}. Moreover, were a loop structure to be
connected to one outer leg representing a tree diagram, the resulting
expression would contain a total derivative ($b$) on the superfields,
from the propagator connecting the tree part to the loop structure.

Consequently, at least two outer legs need to be attached to a loop
cluster in order for the resulting expression to be non-zero. Taking
the possibility of attaching tree diagrams to the loop structure into
consideration, it is also possible to see that two outer legs are
enough for a non-zero result (for some diagrams). Four external fields
can be present and the parts of the propagators that originate outside
of the loop structure split up in a way as to not present a total
derivative on the entire amplitude diagram.

With this, we conclude the section on the description of amplitude
diagrams and move on to their UV properties. Note that there may be
several more features of the amplitude diagrams, not described
above. As previously mentioned though, the structure is complex and
difficult to discern.

\section{Ultraviolet behaviour of amplitude diagrams}\label{sec.UV}
In the presence of loops,
integrations over the free momenta $p^I$ are introduced. These
integrations are performed up to some large momentum cut-off limit
$\La$, which typically threatens to make the expressions
divergent. The infrared behaviour of this needs a detailed
investigation, and will not be discussed here, but the
examination of the ultraviolet behaviour is
rather straightforward. The dependence on the momentum cut-off
goes as:
\begin{equation}\label{eq.UVpower}
\La^{LD-2m+n}
\end{equation}
Here, $L$ is the number of loops, $D$ the dimension (e.g. after a
dimensional reduction), $m$ the number of propagators that are part of
the loops and $n$ the number of loop momenta that can be constructed
and paired up into $(\partial^I)^2$ in the loops. The theory is UV
divergent if \cite{JB}
\begin{equation}
LD-2m+n\geq0
\end{equation}

As discernible above, the dimensional dependence enters through the
integration over the loop momenta $p^I$, which there is $D$ of in each
of the $L$ loops. The second term in the expression ($-2m$) represents
the UV contribution of loop momenta from the propagators:
\begin{equation}
\frac{1}{(p+p^I)^2}\sim\frac{1}{(p^I)^2}(1+O(p/p^I))\li, \quad p^I \gg p
\end{equation}
Moreover, the last term of $n$ must be even since parts that contain
unpaired $\partial^I$'s give a zero contribution at the integration
over $\mathrm{d}p^I$ .

An interesting question is whether or not maximal supergravity
is finite in,
say, four dimensions. In specific, what conditions does the framework
of the formalism presented above set on the theory, in order for it to
be finite? This question is what we address below, first by a look at
what loop structures look like with respect to their UV contribution,
and later on by discerning what parts of them give at hand the
worst dependence on the momentum cut-off. Lastly, we look at what this
worst dependence looks like, for as it turns out, it is generally
provided by the amplitude diagrams that need regularisation with respect to the number of $r$'s.

\subsection{UV characteristics}
The only part of an amplitude diagram which contributes to the UV
behaviour, brought on by the loop regularisation, is the loop
structure. The behaviour is described by eq. \eqref{eq.UVpower}, where
only one part is unknown, the term $n$ describing the
number of loop
momenta that can be paired up into scalars in the loop
structure. Furthermore, the number $n$ is only
unknown for
amplitudes which are regularised with respect to $r$, a procedure
we will refer to as just ``regularisation'' from here on.

Recall, from the previous section, that no quantity $(\partial^I)^2$ can be formed out of unregularised components. For diagrams that do not need to be regularised, we therefore can conclude that the worst UV dependence appears for diagrams only consisting of 3-point vertices, for which the number of outer legs are as few as possible, in the following way:
\begin{equation}\label{eq.UVnoReg}
\La^{L(D-6)+6-2j}
\end{equation}
The reason for the restriction to a consideration of the minimal $j$
is obvious: it is what gives the worst UV behaviour, which is what we
are interested in. Moreover, the demand on $D$ in order for this
exponent not to be negative for any $L$ or $j$ is not more severe than
$D<\frac{11}{2}$. This would arise from $L=4$, $j=2$, as the minimal
$j$'s increase for low $L$'s, if indeed that amplitude does not need
to be regularised. The question which remains to be answered is which
conditions we have in general, as a function of $L$.

In order to determine the worst possible UV behaviour in general, we
need to know when it is necessary to perform regularisation with
respect to $r$, and how to interpret the components of the amplitude
diagrams. This procedure might be anticipated to be tedious, but there
is a convenient simplification for the case when the worst UV
divergence is sought for, as is applicable here. It is possible to use
the correspondence $r\leftrightarrow D$ and
$D^2\leftrightarrow \partial$ in the search for when and how to
regularise, and ultimately in the search of $n$. As thus the problem
is reduced to a matter of a counting of the power of $r$ that is
present, and an examination of whether or not it is allowed to be
formed according to the amplitude characteristics. The problem
separates into two parts. The first one refers to the actual power of
$r$ present, when the conditions for a requirement of regularisation
are to be determined. The second refers to an effective power of $r$:
$r^3/p^2\sim r^{-1}$, which in the end gives at hand the UV
characteristics. Some properties are related though, which is why part
of the discussion here will be performed simultaneously. We begin by
having a look at the different parts of the amplitude diagrams, with
respect to the above.

\subsubsection{The insignificance of components outside the loop}
Parts outside of the loop structure, e.g. external legs or parts of a
tree structure connecting the loop structure to external legs, do not
change the UV behaviour of the loop. They only way they could do so,
by causing a higher power of $r$ to be present and further
regularisations to be necessary, provides only a superficial change,
as noted in subsection \ref{subsub.matterLoop}. As such, we consider only the loop structure in order to determine the
properties of it.

In supergravity, where vertex operators are present, this also means
that operators that are associated with outer vertices can be assumed
to act out of the loop as much as possible. Which leg each of them act
on is arbitrary, yielding the same result, and thus the version where
as many $r$'s as possible get outside of the loop, not to be included
in the loop regularisation, reflects the real properties of the loop
with the least intermediate examination. For example, both 3- and
4-point vertices (with 2 legs as part of the loop structure) can be
taken to contribute with $R^a$ only.

Furthermore, for regularised amplitudes, the number of outer legs is
irrelevant. The conversion of a 3-point vertex into a 4-point one,
with the new leg as an outer leg does not contribute anything new to
the loop structure, as the $T$ can be assumed to act out of the
loop. Moreover, only one extra $R^a$ is contributed to the loop by the
addition of an outer vertex, and the additional $b$ at most can
contribute the equivalent of $r^2$ to the loop (since each loop only
can have one outer propagator completely made out of loop variables),
so that introduction of an outer vertex goes as $r^4/p^2 \sim p^0$. Consequently, for regularised
amplitudes, the number of $j$ is irrelevant.

\subsubsection{The loop structure}
Let us start by counting the maximal power of $r$ that may be present
in a loop structure. Beginning with the $b$ operator, we may note that
its maximal contribution corresponds to $r^3$. Similarly, the $R^a\sim
r^2$ and $T\sim r^3$. Consequently, the 3-point vertex is represented
by $r^4$ and the 4-point vertex by $r^7$. Furthermore, an amplitude
diagram with only 3-point vertices contains $3(L-1)+j$ propagators and
$2(L-1)+j$ vertices, compare subsection \ref{subsub.loopstr}.

The presence of a 4-point vertex decreases the number of propagators
by one, and the number of 3-point vertices by two, giving at hand that
the maximal power of $r$ present, for internal vertices, are
\emph{less} by a factor $r^4$ compared to a diagram of only 3-point
vertices. For outer vertices, the difference is at most $r^{-5}$, and
the presence of the 4-point vertex can be recognised as insignificant
due to the following reasons:

\begin{itemize}
\item{For regularised amplitudes, the effective power of $r$ is unaffected by the 4-point vertex. The presence of $1/p^2\sim r^{-4}$ in the propagator, so the 4-point vertex and the 4-point tree amplitude built from 3-point vertices are indistinguishable in the power counting of $r$'s. Outer vertices do not need to be considered, as mentioned in the previous subsection.}

\item{For low values of $L$, the presence of the 4-point vertex
    increases the minimal number of $j$ to be attached by one. This is
    true for up to four loops, and we will see that higher loops
    require regularisation. However, when $j$ is increased by one, a
    new set of vertex operators is added and the maximal power of $r$
    increased by $r^7$. Effectively, the maximal power of $r$ is thus
    increased by at least $r^2$ by the presence of the 4-point vertex,
    and it need not be considered when the question of when
    regularisation with respect to $r$ is needed is addressed.}
\end{itemize}

In short, the above states that only 3-point vertices need to be
considered. For outer legs, the contribution to the maximal power of
$r$ is $r^5$, and the effective contribution is $r^0$. Apart from
that, the relevant numbers of propagators and inner vertices are
$3(L-1)$ and $2(L-1)$. The contributions with respect to the maximal
number of $r$ for each such component are $r^3$ and $r^4$
respectively.

As such, the maximal power of $r$ present in a loop diagram is given by the term $9(L-1)+8(L-1)+5j$. In order for this to give at hand new results with
respect to the $r\leftrightarrow D$ regularisation, the power must
exceed that of $r^{23}D^{9L}$:
\begin{align}\label{eq.sugraCond}
 8L+5j>40
\end{align}
That is, when the condition above is fulfilled, the $r\leftrightarrow D$
conversion must be taken into account, and eq. \eqref{eq.UVnoReg} no
longer holds. Note that this always is true for \makebox{$L>3$}, part of which
was assumed above when the 4-point amplitude was neglected. Moreover,
it is true for any $L>1$ when the number of outer legs is not
minimal.

\subsubsection{Regularised amplitudes}
In the search of the quantity $n$, the effective power of $r$ present
in a diagram is to be considered. The different parts are listed
above, giving at hand $r$ to the power of $9(L-1)+8(L-1)$. Out of
these, $23+9L$ are claimed by the loop regularisations and the $r$'s
that need to be left untouched. The rest, $8L-40$, may be converted
into $4L-20$ $p$'s. Furthermore, for the worst UV dependence, they are
assumed to pair up into $(p^I)^2$'s to as great an extent as possible,
which all of them can do as the number is even, and no further
restrictions on the amplitude diagrams have been discerned. Thus
$n=4L-20$, giving at hand a maximal dependence on the cut-off
$\Lambda$:
\begin{align}
\La^{LD-2m+n}&=\big[\li m=3(L-1), \quad n=4L-20\li\big]=\nonumber\\
&=\La^{L(D-2)-14}\li,\quad 8L+5j>40
\end{align}
Where this result is not valid, the dependence on $\La$ is governed by
the unregularised result. Moreover, note that the regularised result
for $L=4$, $j=2$ is the same as if it had not been regularised, which
originates in the property that the superfluous $r^2$ cannot
form a term $\partial^2$.

\subsection{The UV dependence}
To conclude, the previous subsection states that the the maximal dependences on $\La$ and the subsequent requirements on $D$ in order for the theory to be UV finite are:
\begin{equation}\begin{array}{llcl}
\La^{D-2j}&L=1, 4\leq j\leq6 &:&D<2j\\
\La^{2D-14}&L=2, j=4 &:&D<7 \\
\La^{3D-18}&L=3, j=3 &:&D<6 \\
\La^{L(D-2)-14}\quad&\text{otherwise} &:&D<2+\frac{14}{L}
\end{array}
\end{equation}
This fits exactly with previous results, for example the ones in
ref. \cite{Z.B} for $L\leq4$, and in refs. \cite{BjornssonGreen,JB}
for all $L$'s, and points towards a possible divergence at $L=7$. 

\section{Conclusions and outlook}\label{sec.Con}
The characteristics of the amplitude diagrams in maximal supergravity are intricate, and so far the description
is not exhaustive. There may yet be further properties to be discovered, that
would affect the convergence
of the diagrams in the UV regime. So far
though, the maximally supersymmetric description with manifest
supersymmetry through the use of pure spinors, has not yielded any
other results than the first quantised version, based on $D=10$
  pure spinors, presented in ref.
\cite{JB}. The predictions contain the possible
divergence of $D=4$, $\mathcal{N}=8$
supergravity at seven loops.

This is to be compared to maximal SYM, which is known to be perturbatively finite in four dimensions, a fact that is predicted by a similar examination as above, but for SYM.  The principles are the same though the degrees of freedom differ\footnote{There are 11 degrees of freedoms for pure
  spinors and otherwise 16 (compare 23 and 32) so that 5 $D$'s are
  required for a loop integration.} and no vertex operators are
present. The result for the unregularised case is the same as in
supergravity, followed by $D<4+\frac{6}{L}$ for $L\geq2$.

Of course,
this does not prevent that other characteristics, not yet observed,
might come into play and soften the UV divergence. The main ingredient
that may impose further restrictions on the behaviour of amplitudes is
U-duality. However, it is not known how to incorporate U-duality in a
formulation with pure spinors, and a classical formulation possessing
both manifest U-duality and supersymmetry is still lacking.
U-duality, in $D=4$ manifested as a continuous $E_{7(7)}$ symmetry of
perturbative amplitudes, has been shown to exclude 4-point
counterterms up to six loops, while it seems to leave room for a
possible divergence at seven loops \cite{BEFKMS}. It is striking that
supersymmetry and U-duality seem to yield the same non-renormalisation
properties, and it would indeed be interesting if they could be combined.

Essentially, the more detailed amplitude
calculations of ref. \cite{Z.B} have not yet surpassed the boundary where
the full properties of the loop regularisation come into play, at five
loops. Once that is the case, it might be possible to predict the true UV divergence with more reliability.

\section*{Acknowledgments}
MC would like to thank N. Berkovits for helpful 
discussions.

\appendix
\section{Spinor and pure spinor identities in $D=11$}\label{app.Fierz}
The spinors of supergravity in $D=11$ are symplectic. A spinor index is raised by the
antisymmetric tensor $\varepsilon^{\alpha\beta}$, the presence
of which is usually left implicit. Moreover, note the convention which
we use for the antisymmetrisation of indices:
\begin{equation}
(\g_{ab})_{\alpha\beta}=\frac{1}{2}\big[(\g_{a}\g_{b})_{\alpha\beta}-(\g_{b}\g_{a})_{\alpha\beta}\big]
\end{equation}

Fierz rearrangements are always made between spinors at the right and left of two spinor products, and the general Fierz identity is:
\begin{equation}
(AB)(CD)=\sum\limits_{p=0}^5
\frac{1}{32p!}(C\g^{a_1\ldots a_p}B)(A\g_{a_p\ldots a_1}D)
\end{equation}
In the above, the spinors have been assumed to be bosonic, but the relation holds for all operators, provided that the appropriate sign (for the statistics of the operators) is added. In specific, for bilinears in a pure spinor $\la$ this reduces to:
\begin{equation}
(A\la)(\la B)=-\frac{1}{64}(\la\g^{ab}\la)(A\g_{ab}B)
+\frac{1}{3840}(\la\g^{abcde}\la)(A\g_{abcde}B)
\end{equation}

This gives at hand some useful identities for pure spinors, among which are:
\begin{equation}
\begin{array}{l}
(\g_j\la)_\alpha(\la\g^{ij}\la)=0\\
(\g_i\la)_\alpha(\la\g^{abcdi}\la)=6(\g^{[ab}\la)_\alpha(\la\g^{cd]}\la)\\
(\g_{ij}\la)_\alpha(\la\g^{abcij}\la)=-18(\g^{[a}\la)_\alpha(\la\g^{bc]}\la)\\
(\g_{ijk}\la)_\alpha(\la\g^{abijk}\la)=-42\la_\alpha(\la\g^{ab}\la)\\
(\g_{ij}\la)_\alpha(\la\g^{abcdij}\la)=-24(\g^{[ab}\la)_\alpha(\la\g^{cd]}\la)\\
(\g_i\la)_\alpha(\la\g^{abcdei}\la)=\la_\alpha(\la\g^{abcde}\la)
-10(\g^{[abc}\la)_\alpha(\la\g^{de]}\la)\\
\end{array}
\end{equation}
Furthermore, it is possible to act on the identity $(\lb\g^{[ij}\lb)(\lb\g^{kl]}\lb)=0$ with $\bar\partial$ and make use of the constraint on the spinor $r$, $(\lb\g^a r)=0$, in order to derive:
\begin{equation}\label{LambdaBarRRelOne}
(\lb\g^{[ij}\lb)(\lb\g^{kl]}r)=0
\end{equation}
Other useful relations, especially related to calculations with structures as in eq. \eqref{eq.L}, are:
\begin{equation}\label{LambdaBarRRelTwo}
\begin{array}{l}
(\lb\g^i{}_k\lb)(\lb\g^{jk}r)=(\lb\g^{ij}\lb)(\lb r)\\
(\lb\g^i{}_kr)(\lb\g^{jk}r)=(\lb\g^{ij}r)(\lb
r)+\frac{1}{2}(\lb\g^{ij}\lb)(rr)\\
(\lb\g^i{}_k\lb)(\lb\g^k{}_lr)(\lb\g^{lj}r)=0\\
(\lb\g^i{}_kr)(\lb\g^k{}_lr)(\lb\g^{lj}r)=0\\
\end{array}
\end{equation}
Further relations can of course be deduced, but the above constitute
the main ones used in this paper.

\section{The zero-mode cohomology of $\psi$}\label{app.zmc}
The condition $(\la D)\psi=0$, the equations of motion for a free
field in the minimal formalism, allows for certain non-zero components
of $\psi$, independent of $x$, a set which is termed to be the
zero-mode cohomology of $\psi$. In specific, the variables
($\la,\theta$) must be put together so that they form special
structures, defined up to some general fields with certain
properties. This is what is listed in table \ref{table.CPsi}.

In specific, what the table shows is the field components that are
independent of ($\la,\theta$), listed with respect to their
irreducible representations at the positions which give at hand their
ghost numbers and dimensions\footnote{Recall that $\la$ is of ghost
  number 1, whereas $\theta$ is of ghost number 0. Both are spinors
  and as such of dimension $-1/2$. The superfield, on the other hand,
  has ghost number 3 and \mbox{dimension $-3$.}}. That is, the element in
the upper left corner represents a field in the irreducible
representation of $(00000)$ with ghost number 3 and dimension $-3$,
which represents cohomology in $\psi$ without being attached to
($\la,\theta$). One step to the right represents allowing for one pure
spinor in the superfield, decreasing the ghost number and the
dimension of a field component to 2 and $-5/2$ respectively, though
that spot is empty, represented by a dot. In order to allow for
zero-mode cohomology, at least one $\theta$ must be added. This brings
us one step further down and allows for a field in the representation
of (10000), connected to $\la\g^a\theta$, and so on.

Even though the zero-mode cohomology, the calculation of which is
  a purely algebraic problem, represents a huge
  simplification compared to the actual cohomology, it has a very
  concrete physical interpretation. It is clear that the reintroduction
of the term in $(\la D)$ which contains a derivative with respect to $x$
will impose a further restriction compared to the zero-mode cohomology. The full cohomology will be represented by elements in the
zero-mode cohomology restricted by some differential equations. These
equations will be the equations of motion of the physical
\emph{fields} (including ghosts and antifields), which
in turn are the elements of the zero-mode cohomology. In order for
such equations of motion to contain non-trivial information, they must
in turn be represented by elements in the zero-mode cohomology, but
(since $\la D$ has ghost number $1$) at the next higher power of
$\la$. It is for this reason that both component fields and antifields
(which of course represent equations of motion) appear in the
cohomology, and the models inherently become meaningful only
in a field--antifield, i.e., Batalin--Vilkovisky framework.

\begin{table}
\renewcommand{\arraystretch}{2}
\renewcommand{\tabcolsep}{0.05cm}
\begin{tabular}{c|ccccccccc}
\backslashbox{\small{dim~}}{\small{gh\#~~}}&\hspace{0.2cm}3&2&1&0&$-1$&$-2$&$-3$&$-4$&$\hspace{0.2cm}-5$\\\hline
$-3$&\hspace{0.2cm}{\footnotesize(00000)}&&&&&&&&\\
$-\frac{5}{2}$&\hspace{0.2cm}$\bullet$&$\bullet$&               &&&       &\\
           $-2$&\hspace{0.2cm}$\bullet$&{\footnotesize(10000)}&$\bullet$&       &&&       &\\
      $-\frac{3}{2}$&\hspace{0.2cm}$\bullet$&$\bullet$&$\bullet$&$\bullet$&&&       &\\
          $-1$&$\hspace{0.2cm}\bullet$&$\bullet$&${\footnotesize\raise5pt\vtop{\baselineskip6pt\ialign{\hfill$#$\hfill\cr
					\ss(01000)\cr
					\ss(10000)\cr}}}$
			&$\bullet$&$\bullet$&&\\
      $-\frac{1}{2}$&\hspace{0.2cm}$\bullet$&$\bullet$&{\footnotesize(00001)}
				&$\bullet$&$\bullet$&$\bullet$&&\\
           0&$\hspace{0.2cm}\bullet$&$\bullet$&$\bullet$&${\footnotesize\raise9pt\vtop{\baselineskip6pt\ialign{
					\hfill$#$\hfill\cr
					\ss(00000)\cr
					\ss(00100)\cr
					\ss(20000)\cr}}}
				$&$\bullet$&$\bullet$&$\bullet$&&\\
       $\frac{1}{2}$&\hspace{0.2cm}$\bullet$&$\bullet$&$\bullet$&${\footnotesize\raise5pt\vtop{\baselineskip6pt\ialign{
					\hfill$#$\hfill\cr
					\ss(00001)\cr
					\ss(10001)\cr}}}
				$&$\bullet$&$\bullet$&$\bullet$&$\bullet$&\\
           1&\hspace{0.2cm}$\bullet$&$\bullet$&$\bullet$&$\bullet$&$\bullet$&$\bullet$&$\bullet$&$\bullet$&\hspace{0.5cm}$\bullet$\\
      $\frac{3}{2}$&\hspace{0.2cm}$\bullet$&$\bullet$&$\bullet$&$\bullet$
				&${\footnotesize\raise5pt\vtop{\baselineskip6pt\ialign{
					\hfill$#$\hfill\cr
					\ss(00001)\cr
					\ss(10001)\cr}}}$
				&$\bullet$&$\bullet$&$\bullet$&\hspace{0.5cm}$\bullet$\\
           2&\hspace{0.2cm}$\bullet$&$\bullet$&$\bullet$&$\bullet$
				&${\footnotesize\raise9pt\vtop{\baselineskip6pt\ialign{
					\hfill$#$\hfill\cr
					\ss(00000)\cr
					\ss(00100)\cr
					\ss(20000)\cr}}}
				$&$\bullet$&$\bullet$&$\bullet$&\hspace{0.5cm}$\bullet$\\
       $\frac{5}{2}$&\hspace{0.2cm}$\bullet$&$\bullet$&$\bullet$&$\bullet$&$\bullet$&{\footnotesize(00001)}&$\bullet$
			&$\bullet$&\hspace{0.5cm}$\bullet$\\
       	   3&\hspace{0.2cm}$\bullet$&$\bullet$&$\bullet$&$\bullet$&$\bullet$
				&${\footnotesize\raise5pt\vtop{\baselineskip6pt\ialign{
					\hfill$#$\hfill\cr
					\ss(01000)\cr
					\ss(10000)\cr}}}$
				&$\bullet$&$\bullet$&\hspace{0.5cm}$\bullet$\\
       $\frac{7}{2}$&\hspace{0.2cm}$\bullet$&$\bullet$&$\bullet$&$\bullet$&$\bullet$&$\bullet$
				&$\bullet$&$\bullet$&\hspace{0.5cm}$\bullet$\\
       	   4&\hspace{0.2cm}$\bullet$&$\bullet$&$\bullet$&$\bullet$&$\bullet$&$\bullet$&{\footnotesize(10000)}
			&$\bullet$&\hspace{0.5cm}$\bullet$\\
       $\frac{9}{2}$&\hspace{0.2cm}$\bullet$&$\bullet$&$\bullet$&$\bullet$&$\bullet$&$\bullet$&$\bullet$
			&$\bullet$&\hspace{0.5cm}$\bullet$\\
       	   5&\hspace{0.2cm}$\bullet$&$\bullet$&$\bullet$&$\bullet$&$\bullet$&$\bullet$&$\bullet$
				&{\footnotesize(00000)}&\hspace{0.5cm}$\bullet$
\end{tabular}
\caption{The zero-mode cohomology in $\psi$. The horizontal direction represents the expansion of the superfield in terms of $\la$ whereas the corresponding for the vertical (in each row) is $\theta$ (downward). The irreducible representations of the component fields are listed at the positions which describe their ghost numbers and dimensions.\label{table.CPsi}}
\end{table}

\section{The projector on the pure spinor cotangent space}\label{app.Pro}
What may be used in order to ensure the properties of the projection $P^\alpha{}_\beta$ in eq. \eqref{eq.projector}, described in subsection \ref{subsec.genTransf}, is  the Fierz identity  $(\g_b\la)_\alpha(\la\g^{ab}\la)=0$. Therefore, we write an ansatz for $P_\perp$:
\begin{equation}
P_\perp^\alpha{}_\beta=(\g_a\fb)^\alpha R^a{}_b(\la\g^b)_\beta \equiv \Pi^\alpha{}_\beta[R]
\end{equation}
Here, $R$ is some, yet undetermined, matrix, $\fb$ is an
arbitrary non-vanishing pure reference spinor, and the last equality
just represents a convenient notation. It is not immediately obvious
how to best make an ansatz for $R$, since there is a number of
independent expressions containing an equal number of $\la$'s and
$\fb$'s. In addition to the unit matrix, there are two antisymmetric
matrix structures and two symmetric. They may be taken as:
\begin{equation}
\begin{array}{ll}
M_{ab}\equiv(\la\g_{ab}\fb) &\qquad S^{(2)}_{ab}\equiv(\la\g_{(a}{}^i\la)(\fb\g_{b)i}\fb)  \\
A^{(2)}_{ab}\equiv
(\la\g_{[a}{}^i\la)(\fb\g_{b]i}\fb)&\qquad S^{(5)}_{ab}\equiv(\la\g_{(a}{}^{ijkl}\la)(\fb\g_{b)ijkl}\fb)
\end{array}
\end{equation}
However, since the matrices appear sandwiched between $\g_a\fb$ and $\la\g^b$, $A^{(2)}$ and $S^{(2)}$ are equivalent: $\Pi[A^{(2)}]=\Pi[S^{(2)}]$. Alternatively, one of the antisymmetric expressions may be replaced by $A^{(5)}_{ab}\equiv(\la\g_{[a}{}^{ijkl}\la)(\fb\g_{b]ijkl}\fb)$, thanks to the Fierz identity $\alpha M_{ab}=\frac{1}{16}A^{(2)}_{ab}-\frac{1}{384}A^{(5)}_{ab}$. There are two scalar invariants, $\alpha=(\la \fb)$ and
$\beta=(\la\g_{ab}\la)(\fb\g^{ab}\fb)$, and the ansatz for the matrix will consist of functions of these invariants multiplying the matrices. More precisely, we will have functions of the dimensionless invariant $x=\frac{\beta}{8\alpha}$, and everywhere the appropriate power of $\alpha$ to make all terms dimensionless. For simplicity we will temporarily set $\alpha=1$ below.

In doing this, it is easier to work with a basis for the matrices consisting of
powers of $M^a{}_b=(\la\g^{a}{}_b\fb)$. This is because the multiplication rule is
\begin{equation}
\Pi[R]\Pi[S]=\Pi[R(\mathbb{I}+M)S]
\end{equation}
which follows immediately from the definition of $\Pi[R]$.
It therefore follows that $\Pi[(\mathbb{I}+M)^\star]$ is a projection
matrix, where the operation $R^\star$ denotes a ``weak inverse'' in the
sense that $\Pi[R^\star R]=\Pi[\mathbb{I}]$. This far, the
  construction is completely
general and independent of dimension. The specific information will
turn up in an effective Cayley--Hamilton equation for the matrix $M$,
which will be used in the form
\begin{equation}\label{PiPM}
\Pi[p(M)]=0
\end{equation}
where $p$ is some polynomial. From the considerations
above, the degree of $p$ must be\footnote{Note that in $D=10$, the degree
is 1, and $\Pi[M-\mathbb{I}]=0$.} 4.
Multiplication by $M$ is easy for $M$, and also for $A^{(2)}$ an
$S^{(2)}$, obeying $MA^{(2)}=S^{(2)}$, $MS^{(2)}=A^{(2)}$. A
calculation with some Fierz rearrangements leads to
\begin{align}
M^3-(1-x)M&=\frac{1}{2}A^{(2)}\\
M^4-(1-x)M^2&=\frac{1}{2}S^{(2)}
\end{align}
and subtraction of these two equations leads to an equivalence of the form in eq. \eqref{PiPM} with
\begin{equation}
p(M)=M^4-M^3-(1-x)M^2+(1-x)M
\end{equation}
Note that the Cayley--Hamilton relation $\Pi[p(M)]=0$ is stronger
than, but consistent with, the one for the matrix $M$ itself:
\begin{equation}
M^5-(2-x)M^3+(1-x)M=0
\end{equation}

The projection matrix can now be calculated using this stronger
Cayley--Hamil\-ton relation, and we find:
\begin{equation}
\Pi[(\mathbb{I}+M)^\star]=\Pi[\mathbb{I}-\frac{1+x}{2x}M+\frac{1}{x}M^2-\frac{1}{2x}M^3]
\end{equation}
This can of course be translated to another basis (the basis
$\{\mathbb{I},M,M^2,M^3\}$ has the disadvantage of containing structures with
unnecessarily high powers of $\la$ and $\fb$), see below.

We can verify that this projection indeed is on the complement to the
tangent space of pure spinor space by calculating its rank. Using
$\tr\Pi[R]=\tr(\mathbb{I}+M)R$, and
\begin{align}
\tr\mathbb{I}&=11\\
\tr M^2&=2(5-3x)\\
\tr M^4&=(1-x)\tr M^2+4x=2(5-6x+3x^2)
\end{align}
we arrive at $\tr\Pi[(\mathbb{I}+M)^\star]=9$, so this is indeed the correct projection.

A gauge invariant derivative can be formed as:
\begin{equation}
W_\alpha=(P^tw)_\alpha
\end{equation}
Its invariance, when ${P^\alpha}_\beta={\delta^\alpha}_\beta-\Pi^\alpha{}_\beta[R]$, means that:
\begin{equation}
(\la\g^a)_\alpha-((\mathbb{I}+M)R)^a{}_b(\la\g^b)_\alpha=0
\end{equation}
This invariance follows directly from the general relations above (in
any dimension), from the observation that the multiplication with $\la\g^b$ on
the right is sufficient to imply the stronger Cayley--Hamilton
relation. This projected derivative can be seen as the covariant
derivative (with $\fb\rightarrow\lb$) corresponding to the metric
obtained as the pullback from flat $32_{\mathbb C}$-dimensional spinor
space.

A nicer basis might be the one based on the matrices $\mathbb{I}$, $M$,
$A^{(2)}\approx S^{(2)}$ and $S^{(5)}-A^{(5)}$. The last matrix is
chosen since its contribution can be rewritten using
\begin{equation}
\Pi[S^{(5)}-A^{(5)}]=
36(\g_{[ij}\fb)^\alpha(\fb\g_{kl]}\fb)(\la\g^{ij}\la)(\la\g^{kl})_\beta
\end{equation}
The expression for the projector to tangent space then is
\begin{align}
P^\alpha{}_\beta&=\delta^\alpha{}_\beta-\frac{1}{4\alpha}(\g_a \fb)^\alpha(\la\g^a)_\beta
-\frac{1}{2\alpha\beta}(\g_a \fb)^\alpha(\la\g^{ai}\la)(\fb\g_{bi}\fb)(\la\g^b)_\beta+\nonumber\\
&\quad
+\frac{3}{4\alpha\beta}(\g_{[ij}\fb)^\alpha(\fb\g_{kl]}\fb)(\la\g^{ij}\la)(\la\g^{kl})_\beta
\end{align}
since the coefficient of $\Pi[M]$ vanishes. Here, the explicit powers of
$\alpha$ have been reinserted. From this form we can more easily verify that
no terms blow up when one approaches the codimension 7 subspace with
$(\la\g_{ab}\la)=0$.

A ``translated'' $\la$ is obtained by exponentiating:
\begin{equation}
\la'=e^{(\e W)}\cdot\la^\alpha=\la^\alpha+\e^\alpha-\frac12(\g_a\fb)^\alpha R^a{}_b(\la+\e,\fb)
((\la+\e)\g^b(\la+\e))
\end{equation}
This shows that a regulator whose action on the bosonic variables is
given by eq. \eqref{eq.GuessOp} provides enough smoothing around
$\eta=0$ to render bosonic integration finite, as described in subsection
\ref{subsec.genTransf}.

\selectlanguage{english}


\begin{thebibliography}{99}
%
\bibitem{BSS}L. Brink, J. Scherk and J. H. Schwarz, \emph{Supersymmetric
      Yang--Mills theories}, Nucl. Phys. {\bf B121} (1977) 77.
%
\bibitem{BLN}L. Brink, O. Lindgren and B. E. W. Nilsson, \emph{The
    ultraviolet finiteness of the $\mathcal{N}=4$ Yang--Mills theory},
  Phys. Lett. {\bf B123} (1983) 323.
%
\bibitem{Mandelstam}S. Mandelstam, \emph{Light cone superspace and the
  ultraviolet finiteness of the $\mathcal{N}$=4 model}, Nucl. Phys. {\bf B213}
(1983) 149.
%
\bibitem{HST}P. S. Howe, K. S. Stelle and P. K. Townsend,
  \emph{Miraculous ultraviolet cancellations in supersymmetry made manifest},
  Nucl. Phys. {\bf B236} (1984) 225.
%
\bibitem{deWitNicolai}B. de Wit and H. Nicolai, \emph{$\mathcal{N}$=8
    supergravity}, Nucl. Phys. {\bf B208} (1982) 323.
%
\bibitem{ElevenSG}E. Cremmer, B. Julia and J. Scherk,
\emph{Supergravity theory in eleven-dimensions},
Phys. Lett. {\bf B76} (1978) 409.
%
\bibitem{ElevenSSSG}L. Brink and P. Howe,
\emph{Eleven-dimensional supergravity on the mass-shell in superspace},
Phys. Lett. {\bf B91} (1980) 384.
%
\bibitem{CremmerFerrara}E. Cremmer and S. Ferrara,
\emph{Formulation of eleven-dimensional supergravity in superspace},
Phys. Lett. {\bf B91} (1980) 61.
%
\bibitem{Kallosh}R. Kallosh (2011) \emph{$E_{7(7)}$ Symmetry and finiteness
    of $\mathcal{N}$=8 supergravity}, [arXiv:1103.4115 [hep-th]].
%
\bibitem{V} P. Vanhove (2010) \emph{The critical ultraviolet behaviour of $\mathcal{N}$=8 supergravity amplitudes} [arXiv:1004.1392 [hep-th]].
%
\bibitem{BjornssonGreen}J. Bj\"ornsson and M. B. Green, \emph{5 loops
    in 24/5 dimensions}, J. High Energy Phys. {\bf1008} (2010) 132
  [arXiv:1004.2692 [hep-th]].
%
\bibitem{JB} J. Bj\"ornsson, \emph{Multi-loop amplitudes in maximally supersymmetric pure spinor field theory}, J. High Energy Phys. {\bf 1101} (2011) 002 [arXiv:1009.5906 [hep-th]].
%
\bibitem{BEFKMS}N. Beisert, H. Elvang, D. Z. Freedman, M. Kiermaier,
  A. Morales and S. Stieberger, \emph{$E_{7(7)}$ constraints on
    counterterms in $\mathcal{N}$=8 supergravity}, Phys. Lett. {\bf B694} (2010)
  265 [arXiv:1009.1643 [hep-th]].
%
\bibitem{BossardHoweStelle}G. Bossard, P. S. Howe and K. S. Stelle,
  \emph{The ultra-violet question in maximally supersymmetric field
    theories}, Gen. Rel. Grav. {\bf41} (2009) 919 [arXiv:0901.4661 [hep-th]].
%
\bibitem{GSB} M. B. Green, J. H. Schwarz and L. Brink, \emph{$\mathcal{N}$=4 Yang--Mills and $\mathcal{N}$=8 supergravity as limits of string theories}, Nucl. Phys. B {\bf 198} (1982) 474.
%
\bibitem{BDDPR} Z. Bern, L. J. Dixon, D. C. Dunbar, M. Perelstein and J. S. Rozowsky, \emph{On the relationship between Yang--Mills theory and gravity and its implication for ultraviolet divergences}, Nucl. Phys. B {\bf 530} (1998) 401 \makebox{[arXiv:hep-th/9802162]}.
%
\bibitem{BCDJKR} Z. Bern, J. J. Carrasco, L. J. Dixon, H. Johansson and R. Roiban, \emph{Three-loop superfiniteness of $\mathcal{N}$=8 supergravity}, Phys. Rev. Lett. {\bf 98} (2007) 161303 [arXiv:hep-th/0702112].
%
\bibitem{Z.B} Z. Bern, J. J. Carrasco, L. J. Dixon, H. Johansson and R. Roiban, \emph{The ultra\-violet behaviour of $\mathcal{N}$=8 supergravity at four loops}, Phys. Rev. Lett. {\bf 103} (2009) 081301 [arXiv:0905.2326v2 [hep-th]].
%
\bibitem{N} B. E. W. Nilsson, \emph{Pure spinors as auxiliary fields in the ten-dimensional supersymmetric Yang--Mills theory}, Class. Quantum Grav. {\bf 3} (1986) L41.
%
\bibitem{H1} P. S. Howe, \emph{Pure spinor lines in superspace and ten-dimensional super\-symmetric theories}, Phys. Lett. {\bf B258} (1991) 141.
%
\bibitem{H2} P. S. Howe, \emph{Pure spinors, function superspaces and supergravity theories in ten and eleven dimensions}, Phys. Lett. {\bf B273} (1991) 90.
%
\bibitem{B1} N. Berkovits, \emph{Super-Poincaré covariant quantization of the superstring}, J. High Energy Phys. {\bf 0004} (2000) 018 [arXiv:hep-th/0001035].
%
\bibitem{B2} N. Berkovits, \emph{Covariant quantization of the superparticle using pure spinors}, J. High Energy Phys. {\bf 0109} (2001) 016 [arXiv:hep-th/0105050].
%
\bibitem{CGNT}M. Cederwall, U. Gran, B.E.W. Nilsson and D. Tsimpis,
\emph{Supersymmetric corrections to eleven-dimen\-sional supergravity},
J. High Energy Phys. {\bf 0505} (2005) 052 [arXiv:hep-th/0409107].
%
\bibitem{CGNN}M. Cederwall, U. Gran, M. Nielsen and B. E. W. Nilsson,
\emph{Manifestly supersymmetric M-theory},
J. High Energy Phys. {\bf 0010} (2000) 041 [arXiv:hep-th/0007035];
\emph{Generalised 11-dimensional supergravity}, [arXiv:hep-th/0010042].
%
\bibitem{CNT1} M. Cederwall, B. E. W. Nilsson and D. Tsimpis, \emph{The structure of maximally supersymmetric Yang--Mills theory: constraining higher-order corrections}, J. High Energy Phys. {\bf 0106} (2001) 034 [arXiv:hep-th/0102009].
%
\bibitem{CNT2} M. Cederwall, B. E. W. Nilsson and D. Tsimpis, \emph{D=10 super-Yang--Mills at O($\alpha'^2$)}, J. High Energy Phys. {\bf 0107} (2001) 042 [arXiv:hep-th/0104236].
%
\bibitem{HoweTsimpis}P. S. Howe and D. Tsimpis, \emph{On higher order
    corrections in M theory}, J. High Energy Phys. {\bf0309} (2003)
  038 [arXiv:hep-th/0305129].
%
\bibitem{CNT3} M. Cederwall, B. E. W. Nilsson and D. Tsimpis, \emph{Spinorial cohomology and maximally supersymmetric theories}, J. High Energy Phys. {\bf 0202} (2002) 009 [arXiv:hep-th/0110069]; M. Cederwall, \emph{Superspace methods in string theory, supergravity and gauge theory}, Lectures at the XXXVII Winter School in Theoretical Physics ``New Developments in Fundamental Interaction Theories'', Karpacz, Poland, Feb. 6-15, 2001 [arXiv:hep-th/0105176].
%
\bibitem{MS} M. Movshev and A. Schwarz, \emph{On maximally supersymmetric Yang--Mills theories}, Nucl. Phys. {\bf B681} (2004) 324 [arXiv:hep-th/0311132].
%
\bibitem{CederwallBLG}M. Cederwall, \emph{$\mathcal{N}$=8 superfield formulation of
the Bagger--Lambert--Gustavsson model}, J. High Energy Phys. {\bf
0809} (2008) 116
[arXiv:0808.3242 [hep-th]].
%
\bibitem{CederwallABJM}M. Cederwall, \emph{Superfield actions for $\mathcal{N}$=8
and $\mathcal{N}$=6 conformal theories in three dimensions},
J. High Energy Phys. {\bf0810} (2008) 070 [arXiv:0809.0318 [hep-th]].
%
\bibitem{C1} M. Cederwall, \emph{Towards a manifestly supersymmetric action for 11-dimensional supergravity}, J. High Energy Phys. {\bf 1001} (2010) 117 [arXiv:0912.1814 [hep-th]].
%
\bibitem{C2} M. Cederwall, \emph{D=11 supergravity with manifest supersymmetry}, Mod. Phys. Lett. {\bf A25} (2010) 3201 [arXiv:1001.0112 [hep-th]].
%
\bibitem{CederwallReview}M. Cederwall, \emph{Pure spinor superfields,
    with applications to D=3 conformal models}, Proc. Estonian
  Acad. Sci. {\bf4} (2010) [arXiv:0906.5490 [hep-th]]; \emph{From supergeometry
  to pure spinors}, [arXiv:1012.3334 [hep-th]].
%
\bibitem{CK} M. Cederwall and A. Karlsson, \emph{Pure spinor superfields and Born--Infeld theory}, J. High Energy Phys. {\bf 1111} (2011) 134 [arXiv:1109.0809 [hep-th]].
%
\bibitem{J1} N. E. J. Bjerrum-Bohr and P. Vanhove, \emph{Explicit cancellation of triangles in one-loop
gravity amplitudes}, J. High Energy Phys. {\bf 0804} (2008) 065 [arXiv:0802.0868 [hep-th]].
%
\bibitem{J2} N. E. J. Bjerrum-Bohr and P. Vanhove, \emph{Absence of triangles in maximal supergravity
amplitudes}, J. High Energy Phys. {\bf 0810} (2008) 006 [arXiv:0805.3682 [hep-th]].
%
\bibitem{KLT}H. Kawai, D. C. Lewellen and S. H. Tye, \emph{A relation
    between tree amplitudes of closed and open strings},
  Nucl. Phys. {\bf B269} (1986) 1.
%
\bibitem{Berkovits&Nekrasov} N. Berkovits and N. Nekrasov, \emph{Multiloop superstring amplitudes from non-minimal pure spinor formalism}, J. High Energy Phys. {\bf 0612} (2006) 029 [arXiv:hep-th/0609012].
%
\bibitem{AB} Y. Aisaka and N. Berkovits, \emph{Pure spinor vertex operators in Siegel gauge and loop amplitude regularization}, J. High Energy Phys. {\bf 0907} (2009) 062 [arXiv:0903.3443 [hep-th]].
%
\bibitem{GV} P. A. Grassi and P. Vanhove, \emph{Higher-loop amplitudes in the non-minimal pure spinor formalism}, J. High Energy Phys. {\bf0905} (2009) 089 [arXiv:0903.3903 [hep-th]].
%
\bibitem{DragonWindow} M. Cederwall, U. Gran and B. E. W. Nilsson, \emph{D=3, $\mathcal{N}$=8 conformal super\-gravity and the Dragon window}, J. High Energy Phys. {\bf 1109} (2011) 101 [arXiv:1103.4530 [hep-th]].
%
\bibitem{AABN} Y. Aisaka, E. A. Arroyo, N. Berkovits and N. Nekrasov, \emph{Pure spinor partition function and the massive superstring spectrum}, J. High Engery Phys. {\bf } 0808(2008) 050 [arXiv:0806.0584 [hep-th]].
%
\bibitem{BV} I. A. Batalin and G. A. Vilkovisky, \emph{Gauge algebra and quantization}, Phys. Lett. {\bf B102} (1981) 27.
%
\bibitem{FHM} A. Fuster, M. Henneaux and A. Maas, \emph{BRST-antifield
    quantization: a short review}, Int. J. Geom. Meth. Mod. Phys. 2
  (2005) 939 [arXiv:hep-th/0506098].
%
\bibitem{B3} N. Berkovits, \emph{Pure spinor formalism as an $\mathcal{N}$=2 topological string}, J. High Energy Phys. {\bf 0510} (2005) 089 [arXiv:hep-th/0509120].
%
\bibitem{CederwallGeometry}M. Cederwall, \emph{The geometry of
pure spinor space}, J. High Energy Phys. {\bf 1201} (2012) 150
[arXiv:1111.1932 [hep-th]].
%
\bibitem{BerkovitsMinimalLoop}N. Berkovits, \emph{Multiloop amplitudes
    and vanishing theorems using the pure spinor formalism for the
    superstring}, J. High Energy Phys. {\bf0409} (2004) 047
  [arXiv:hep-th/0406055].
%
\bibitem{BedoyaGomez}O. A. Bedoya and H. Gomez, \emph{A new proposal
    for the picture changing operators in the minimal pure spinor
    formalism}, J. High Energy Phys. {\bf1108} (2011) 025 [arXiv:1106.1253 [hep-th]].
%
\bibitem{BerkovitsMembrane}N. Berkovits, \emph{Towards covariant
    quantization of the supermembrane}, J. High Energy Phys. {\bf0209}
  (2002) 051 [arXiv:hep-th/0201151].
%
\bibitem{AGV} L. Anguelova, P. A. Grassi and P. Vanhove, \emph{Covariant one-loop amplitudes in D=11}, Nucl. Phys. B {\bf702} (2004) 269-306 [arXiv:hep-th/0408171].
%
\bibitem{HT} M. Henneaux and C. Teitelboim, 1992. \emph{Quantization of gauge systems}, Princeton, New Jersey 08540, USA: Princeton University Press.
%
\bibitem{Siegel} W. Siegel (1988) \emph{Introduction to string field theory},
  [arXiv:hep-th/0107094].
%
\bibitem{CederwallOperators}M. Cederwall, \emph{Operators on pure spinor
space}, AIP Conf. Proc. {\bf1243} (2010) 51.
%
\bibitem{NathanPrivate}{N. Berkovits, private communication.}
%
\bibitem{UnP} Unpublished work by Y. Aisaka and M. Cederwall (2010).
%
\bibitem{MovshevEleven} M. V. Movshev (2011) \emph{Geometry of a desingularization of eleven-dimensional gravitational spinors} [arXiv:1105.0127 [hep-th]].
%
\bibitem{CederwallNilssonSix}M. Cederwall and B. E. W. Nilsson, \emph{Pure
spinors and D=6 super-Yang--Mills}, [arXiv:0801.1428 [hep-th]].
%
\bibitem{Boulanger}N. Boulanger, T. Damour, L. Gualtieri and
  M. Henneaux, \emph{Inconsistency of interacting, multi-graviton
    theories}, Nucl. Phys. {\bf B597} (2001) 127 [arXiv:hep-th/0007220].
%
\bibitem{SpradlinVolovich}M. Spradlin and A. Volovich,
  \emph{Noncommutative solitons on K\"ahler manifolds}, J. High Energy
  Phys. {\bf0203} (2002) 011 [arXiv:hep-th/0106180].
%
\bibitem{J3}N. Arkani-Hamed, F. Cachazo and J. Kaplan, \emph{What is the simplest quantum field
theory?}, J. High Energy Phys. {\bf 1009} (2010) 016 [arXiv:0808.1446
\mbox{[hep-th]}].


\end{thebibliography}
\end{document}